\documentclass[a4paper,usenatbib]{mnras}
\makeatletter

\newcommand{\Rmnum}[1]{\expandafter\@slowromancap\romannumeral #1@}

\newcommand{\gsim}{\lower0.6ex\vbox{\hbox{$\buildrel{\textstyle >}\over{\sim}\ $}}}

\makeatother

\usepackage{newtxtext,newtxmath}
\usepackage{graphicx}	% Including figure files
\usepackage{amsmath}	% Advanced maths commands
\usepackage{subcaption}
\usepackage{booktabs}
\usepackage[dvipsnames,svgnames]{xcolor}
\usepackage{color}

\def\hkpc{h^{-1}\, {\rm kpc}}

\def\hmsun{{h^{-1} M_{\odot}}}
\def\msun{\, M_{\odot}}

\title[Dual and Offset AGN in ASTRID]
{Properties and Evolution of Dual and Offset AGN in the ASTRID Simulation at $z \sim 2$}

\author[N.~Chen et al.]{
Nianyi Chen,$^{1}$\thanks{E-mail: nianyic@andrew.cmu.edu}
Tiziana Di Matteo,$^{1,2}$
Yueying Ni,$^{1,2}$
Michael Tremmel,$^{3}$
Colin DeGraf,$^{4}$
Yue Shen,$^{5,6}$
\newauthor
A.~Miguel Holgado,$^{1}$
Simeon Bird,$^{7}$
Rupert Croft,$^{1,2}$
Yu Feng $^{8}$\\
$^{1}$ McWilliams Center for Cosmology, Department of Physics, Carnegie Mellon University, Pittsburgh, PA 15213 \\
$^{2}$ NSF AI Planning Institute for Physics of the Future, 
Carnegie Mellon  University, Pittsburgh, PA 15213, USA \\
$^{3}$ Astronomy Department, Yale University, P.O. Box 208120, New Haven, CT 06520, USA \\
$^{4}$ Department of Physics, Truman State University, Kirksville, MO 63501, USA \\
$^{5}$ Department of Astronomy, University of Illinois at Urbana-Champaign, Urbana, IL 61801, USA \\
$^{6}$ National Center for Supercomputing Applications, University of Illinois at Urbana-Champaign, Urbana, IL 61801, USA \\
$^{7}$ Department of Physics \& Astronomy, University of California, Riverside, 900 University Ave., Riverside, CA 92521, USA\\
$^{8}$ Berkeley Center for Cosmological Physics and Department of Physics, University of California, Berkeley, CA 94720, USA
}

\date{Accepted XXX. Received YYY; in original form ZZZ}

\pubyear{2021}
\begin{document}
\maketitle

\begin{abstract}
We examine the dual (both BHs active) and offset (one BH active) AGN population (comprising $\sim$ 2000 pairs at $0.5\,\text{kpc}\lesssim\Delta r<30\,\text{kpc}$) at $z=2\sim3$ in the ASTRID simulation covering (360 cMpc)${^3}$. 
The dual (offset) AGN make up $3.0(2.2)\%$ of all AGN at $z=2$. The dual fraction is roughly constant while the offset fraction increases by a factor of ten from $z=4\sim2$. 
Compared with the full AGN population, duals are characterized by a low $M_\text{BH}/M_*$ ratio, a high specific star-formation rate (sSFR) of $\sim 1\,\text{Gyr}^{-1}$, and a high Eddington ratio ($\sim 0.05$, double that of single AGN).
The dual AGN are formed in major galaxy mergers (typically involving $M_\text{halo}<10^{13}\,M_\odot$), with BHs that have similar masses. 
At small separations (when their host galaxies are in the late phase of the merger) duals become $2\sim8$ times brighter (albeit more obscured) than at larger separations. $80\%$ of these bright, close duals merge in the simulation within $\sim500\,\text{Myrs}$.
Notably, the initially less-massive BH in duals frequently becomes the brighter AGN during the galaxy merger.
In offset AGN, the active BH is typically $\gtrsim 10$ times more massive than its non-active counterpart and than most BHs in duals. Offsets are predominantly formed in minor galaxy mergers with the active BH residing in the center of massive halos ($ M_\text{ halo}\sim 10^{13-14}\,M_\odot$). 
In these deep potentials, gas stripping is common and the secondary quickly deactivates.
The stripping also leads to inefficient orbital decay amongst offsets, which stall at $\Delta r\sim5\,\text{kpc}$ for a few hundred Myrs.

% , whereas offset pairs make up $>80\%$ of the inactive but large MBHs ($M_{\rm BH}>10^7\,M_\odot$).
%selected with the canonical $L_\text{bol,12}>10^{43}\,erg/s$ and $M_\text{BH,12}>10^7\,M_\odot$ thresholds ($L_\text{bol,2}<10^{43}\,erg/s$ for offsets), and resolved down to $\Delta r\sim0.5\,\text{kpc}$ .

\end{abstract}


\section{Introduction}

Super-Massive Black Holes (SMBHs) are believed to reside in the center of most massive galaxies \citep[e.g.][]{Kormendy2013}.
As a consequence of the hierarchical structure formation \citep[e.g.][]{Blumenthal1984}, pairs of SMBHs were found in the merger remnant after mergers between two galaxies.
These SMBH pairs slowly spiral toward the center of mass of the newly merged system and remain at a separation of $0.1\sim 100\,{\rm kpc}$ for a few hundred Myrs \citep[e.g.][]{Begelman1980,Milosavljevic2001}, before dynamical friction drives them into a sub-parsec gravitationally bound binary.

During the galaxy mergers, active galactic nuclei (AGN) can be triggered by the gas driven towards the center of the merger remnant and onto the SMBHs \citep[e.g.][]{DiMatteo2005,Hopkins2008}, making these SMBH pairs observable as either dual AGN \citep[when both of the SMBHs are active, e.g.][]{Gerke2007,Comerford2009}, or offset AGN \citep[when only one of the SMBHs is active, e.g.][]{Steinborn2016}.
Because of the tight connections between the galaxy assemblies and SMBH pairs, the detection and characterization of dual and offset AGN are fundamental for understanding the formation and accretion history of SMBHs across cosmic ages.

There have been significant observational efforts to search for these SMBH pairs using various techniques \citep[see e.g.][ for a comprehensive review of recent observational works]{DeRosa2019}. 
Candidates of dual AGN can be found by searching for double-peaked narrow AGN emission lines in optical spectroscopy \citep[e.g.][]{Comerford2009,Barrows2013}, with follow-up confirmation through other bands \citep[e.g.][]{McGurk2011,Shen2011,Fu2012}.
Hard X-ray observations are widely used to detect multiple AGN in a galaxy especially at high redshifts, being less affected by contamination from stellar processes and absorption \citep[e.g.][]{Fragos2013,Lehmer2016}.
Among the observed samples, some controversial conclusions arise likely due to the different selection functions from different observational techniques.
For example, a number of studies find a higher fraction of dual AGN in galaxies with a closer separation, suggesting that galaxy interactions play a role in the AGN triggering process \citep[e.g.][]{Ellison2011,Silverman2011,Liu2012,Koss2012,Satyapal2014,Kocevski2015,Koss2018}.
On the other hand, there are also studies showing no enhanced AGN activity in mergers compared to a matched control sample of inactive galaxies \citep[e.g.][]{Cisternas2011,Mechtley2016}. 

Despite the massive effort in catching AGN in their dual phase, there have been very limited number of $z\gtrsim2$, close separation ($\Delta r\sim {\rm kpc}$) pairs, due to the limitation in spatial resolution to distinguish between the close pairs.
However, very recently, several groups have been pushing the limit of detecting these high-redshift close pairs using novel observational techniques.
For example, \citet{ChenYuChing2022} uses \texttt{varstrometry} with Gaia DR2 \citep[also see e.g.][]{Shen2019,Hwang2020,Shen2021} to identify several $z\gtrsim2$ dual/offset AGN candidates.
\citet{Silverman2020} uses the double quasar samples from the Hyper Suprime-Cam (HSC) Subaru Strategic Program and identified 421 dual AGN candidates out to a redshift of 4.5.
By looking for distinguished stellar bulges in a sample of AGN host galaxies, \citet{Stemo2021} put up a catalog of 204 offset and dual AGN candidates down to a separation of $<4\,{\rm kpc}$, among which a few are $z\gtrsim2$ AGN.
In a very recent work, \citet{Shen2022} characterizes the statistical properties of galactic-scale quasar pairs using a statistically large sample of 60 double quasars.

In light of these recent observations of high redshift AGN pairs, a sample of simulated counterparts is needed to understand the observed sample and its astrophysical implications.
In the realm of idealized galaxy-merger simulation, \citet{Wassenhove2012,Blecha2013,Capelo2017} AGN activation at various pair separations as well as the impact of the galaxy merger parameters such as the host galaxy mass ratio and morphology.
Recent developments in cosmological hydrodynamical simulations also allow studies of dual and offset AGN and galaxy mergers in a cosmological context \citep[e.g.][]{Volonteri2016,Steinborn2016,Tremmel2017,Rosas-Guevara2019,Ricarte2021,Volonteri2021}, where the number of dual AGN relative to all AGN can be calculated at different redshifts.

Among the cosmological simulations mentioned above, very few were able to produce a statistically large sample of kpc-separation AGN pairs at $z\gtrsim2$, due to several reasons.
First, because the dual and offset AGN only make up a few percent of the total AGN population \citep[e.g.][]{Liu2011,Fu2011}, and because bright AGN are already rare at high redshifts, a large $(\gtrsim 100\,{\rm Mpc/h})^3$ cosmological volume is required to produce those pairs.
Moreover, $\sim {\rm kpc}$ spatial resolution is needed in order to resolve pairs separated by a few kpc.
Finally, even for simulations satisfying the above resolution requirements, in most cosmological simulations BHs are pinned to the gravitational potential minimum to avoid artificial kicks of the BH. 
Consequently, during a galaxy merger, the two central MBHs merge too quickly to be captured at the $\sim {\rm kpc}$ separation.
The BH dynamics modeling after the host galaxy merger is even more important for studying offset pairs \citep[e.g.][]{Barth2008,Comerford2012,Comerford2015,Muller2015,Allen2015} , which are thought to originate mostly from galaxy merger events.

The \texttt{Astrid} simulation uniquely meets the above requirements for studying high-redshift AGN pairs \citep{Bird2021, Ni2021, ChenNianyi2022}.
First, with a volume of $(250\,{\rm Mpc}/h)^3$, \texttt{Astrid} contains $>10^4$ massive AGN already at $z=2\sim3$, among which $\gtrsim 3\%$ are in pairs.
More importantly, the high spatial resolution of $\sim 1.5\,{\rm ckpc}/h$ relative to the volume can resolve AGN pairs at close separations a few hundred Myrs after the host galaxy mergers.
Finally, the dynamical-friction modeling  in \texttt{Astrid} allows for one of the first studies of the evolution of $\Delta r \lesssim 1\,{\rm kpc}$ AGN pairs and their activation in the context of cosmological simulations (previously only done in idealized galaxy merger simulations).

This paper is organized as follows: in Section \ref{sec:sim} we introduce the \texttt{Astrid} simulation, in particular the MBH modeling, and describe our selection criterion for the dual and offset AGN from the simulation;
in Section \ref{sec:z=3}, we focus on a sample of dual and offset AGN at $z=2$, and investigate their properties such as the separation, mass/luminosity distributions, host galaxy mass, AGN activation levels and obscuration, with comparisons with high-redshift observations where possible;
then, in Section \ref{sec:evolution}, we characterize the evolution of AGN pairs at $z=3$ during and after the host galaxy merger, with an emphasis on the effect of pericentric passages and the difference between the evolution of dual and offset AGN.

\section{Simulation}
\label{sec:sim}
The \texttt{Astrid} simulation is a large-scale cosmological hydrodynamic simulation in a $250\, {\rm Mpc}/h$ box with $2\times 5500^3$ particles. 
\texttt{Astrid} contains a statistical sample of halos which can be compared to future survey data from JWST, while resolving galactic halos down to $10^9 \msun$ (corresponding to 200 dark matter particles). 
The initial conditions are set at $z=99$ and the current final redshift is $z=2$. 
The cosmological parameters used are from \cite{Planck}, with $\Omega_0=0.3089$, $\Omega_\Lambda=0.6911$, $\Omega_{\rm b}=0.0486$, $\sigma_8=0.82$, $h=0.6774$, $A_s = 2.142 \times 10^{-9}$, $n_s=0.9667$. 
The mass resolution of \texttt{Astrid} is $M_{\rm DM} = 6.74 \times 10^6 \hmsun$ and $M_{\rm gas} = 1.27 \times 10^6 \hmsun$ in the initial conditions. 
The gravitational softening length is $\epsilon_{\rm g} = 1.5 \hkpc$ for both DM and gas particles.

\begin{figure*}
\centering
  \includegraphics[width=0.99\textwidth]{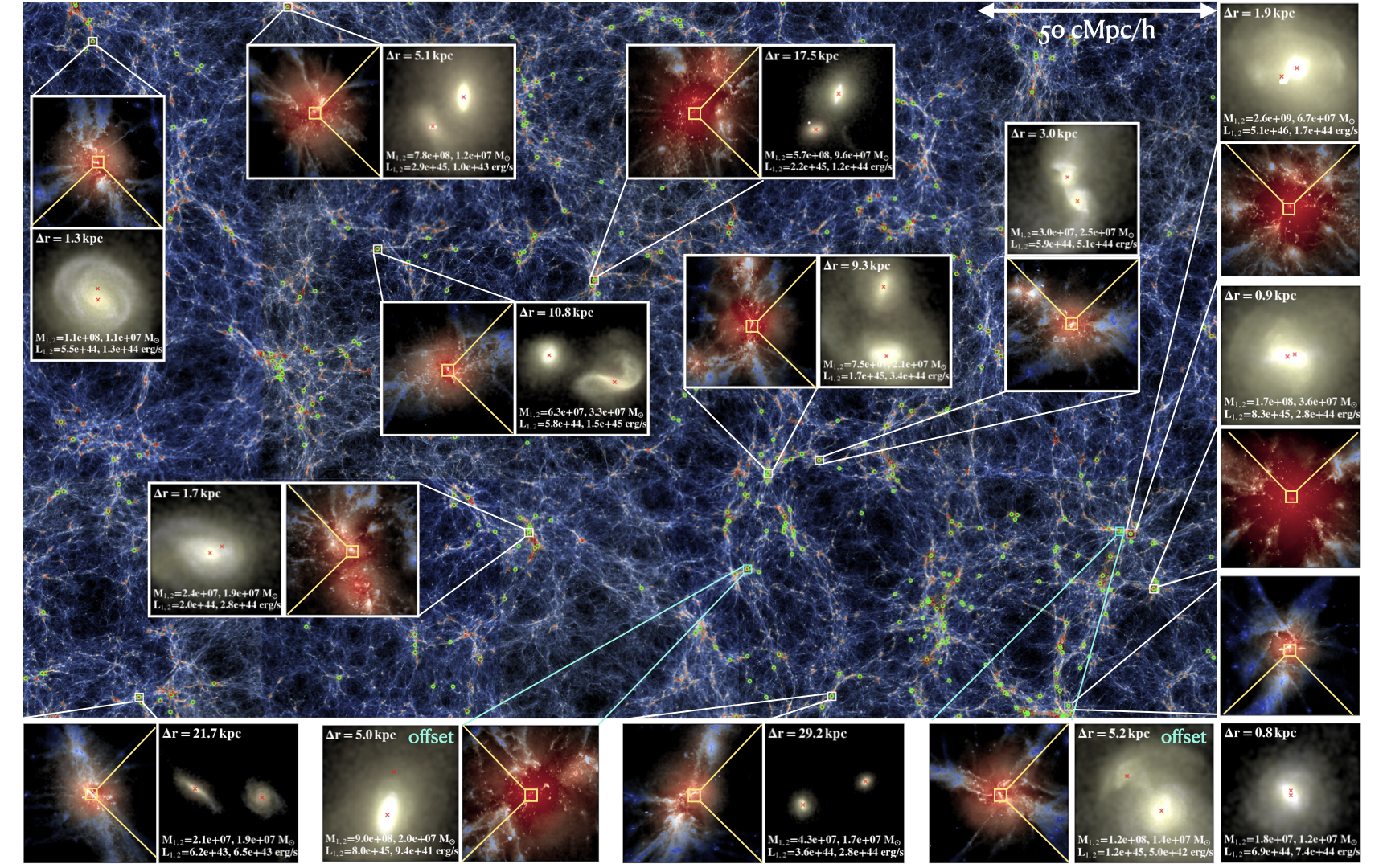}
  \caption{Dual AGN in a $250\,{\rm cMpc}/h\times 150\,{\rm cMpc}/h \times 20\,{\rm cMpc}/h$ slice of the \texttt{Astrid} simulation. The background image shows the gas distribution of the simulation, color-coded by the gas temperature where warmer regions corresponds to higher temperatures. 
  For each dual AGN in the slice, we locate it with white squares in the snapshot, and zoom in to their surrounding IGM and host galaxies. About half of the duals are in separated galaxies with $\delta r > 10 {\rm kpc}/h$, and the other half already in the galaxy merging process.
  }
  \label{fig:img}
\end{figure*}

\subsection{Black Hole Modeling}
Here we briefly describe the BH modeling used in \texttt{Astrid} most relevant for the dual and offset AGN. 
For a thorough description of the sub-grid models and BH statistics, please refer to \citet{Bird2021}, \citet{Ni2021}, and \citet{ChenNianyi2022}.

\texttt{Astrid} contains models for inhomogeneous hydrogen and helium reionization, baryon relative velocities and massive neutrinos, as well as 'full-physics' galaxy formation models including star formation, BH accretion and the associated supernova and AGN feedback. 
The star formation model is unchanged from \cite{Feng2016}, which followed the implementation of \cite{SH03}.
The BH model includes mergers driven by dynamic friction rather than repositioning. 
Our treatment of BHs largely follows the \texttt{BlueTides} simulation in terms of the BH accretion and feedback, which is based on the earlier work by \cite{DiMatteo2005,Springel2005b}.
The gas accretion rate onto the BH is estimated via the Bondi-Hoyle-Lyttleton-like prescription applied to the smoothed properties of the $112$ gas particles within the SPH kernel of the BH.
We allow for short periods of super-Eddington accretion in the simulation, but limit the accretion rate to $2$ times the Eddington accretion rate.
% The MBH produces thermal feedback on the surrounding gas, and radiates with a bolometric luminosity $L_{\rm Bol}$ proportional to the accretion rate $\dot{M}_{\rm BH}$, with a mass-to-light conversion efficiency $\eta=0.1$ in an accretion disk according to \cite{Shakura1973}.
% 5\% of the radiated energy is thermally coupled to the surrounding gas, residing within twice the radius of the SPH smoothing kernel of the BH particle.

The BH radiates with a bolometric luminosity $L_{\rm bol}$ proportional to the accretion rate $\dot{M}_{\rm BH}$, with a mass-to-light conversion efficiency $\eta=0.1$ in an accretion disk according to \cite{Shakura1973}.

We include both thermal (or quasar-mode) feedback and kinetic AGN feedback.
In quasar mode feedback, 5\% of the radiated energy is thermally coupled to the gas residing within twice the radius of the SPH smoothing kernel of the BH particle. A BH switches to the kinetic mode only when the accretion rate drops below the Eddington ratio $\chi_{\text{thr,max}} = 0.05$ and the BH mass is $M_{\rm BH} \gtrsim 10^{8.5} M_\odot$. 
%The Eddington threshold $\chi_{\rm thr}$ for the kinetic feedback is capped at $\chi_{\text{thr,max}} = 0.05$ and is also a function of the BH mass such that the kinetic mode is turned on only for massive black holes with $\mbh \gtrsim 10^{8.5} \msun$. 
The kinetic feedback follows \citet{Weinberger2017}, with slightly different parameters.
Kinetic feedback energy is deposited as $\Delta \dot{E}_{\text {kin}} = \epsilon_{\mathrm{f, kin}} \dot{M}_{\mathrm{BH}} c^{2}$,
where $\epsilon_{\mathrm{f, kin}}$ scales with the BH local gas density and has a maximum value of $\epsilon_{\mathrm{f, kin,max}}=0.05$. 
The energy is accumulated over time and released in a bursty way once the accumulated kinetic feedback energy exceeds the threshold $E_{\mathrm{inj}, \mathrm{min}}=f_{\mathrm{re}} \frac{1}{2} \sigma_{\mathrm{DM}}^{2} m_{\mathrm{enc}}$. 
$\sigma_{\mathrm{DM}}^{2}$ is the one-dimensional dark matter velocity dispersion around the BH, $m_{\mathrm{enc}}$ is the gas mass in the feedback sphere and $f_{\mathrm{re}}=5$.
The released kinetic energy kicks each gas particle in the feedback kernel in a random direction with a prescribed momentum weighted by the SPH kernel.
Kinetic feedback is enabled in \texttt{Astrid} at $z < 2.4$.

To seed MBHs in the simulation, we periodically run a FOF group finder on the fly with a linking length of 0.2 times the mean particle separation, to identify halos with a total mass and stellar mass satisfying the seeding criteria \{ $M_{\rm halo,FOF} > M_{\rm halo,thr}$; $M_{\rm *,FOF} > M_{\rm *,thr}$\}.
We apply a mass threshold value of $M_{\rm halo,thr} = 5 \times 10^9 \hmsun$ and $M_{\rm *,thr} = 2 \times 10^6 \hmsun$.

Instead of applying a uniform seed mass for all BHs, we probe a mass range of the BH seed mass $M_{\rm seed}$ drawn probabilistically from a power-law distribution:
\begin{equation}
\label{equation:power-law}
    P(M_{\rm seed}) = 
    \begin{cases}
    0 & M_{\rm seed} < M_{\rm seed,min} \\
    \mathcal{N} (M_{\rm seed})^{-n} & M_{\rm seed,min} \leq M_{\rm seed} \leq M_{\rm seed,max} \\
    0 & M_{\rm seed} > M_{\rm seed,max}
   \end{cases}
\end{equation}
where $\mathcal{N}$ is the normalization factor.
The minimum seed mass is $M_{\rm seed,min} = 3 \times 10^4 \hmsun$ and the maximum seed mass is $M_{\rm seed,max} = 3 \times 10^5 \hmsun$, with a power-law index $n = -1$. 
For each halo that satisfies the seeding criteria but does not already contain at least one BH particle, we convert the densest gas particle into a BH particle.

Instead of repositioning the black hole towards the potential minimum, in \citet{Chen2021} we implemented and tested a model for sub-grid dynamical friction \citep[similar to][]{Tremmel2015, Tremmel2017}. We set the merging distance to be $2\epsilon_{\rm g}=3\,{\rm ckpc}/h$, because the BH dynamics below this distance is not well resolved. 
We conserve the total momentum of the binary during the merger.
Moreover, when we turn off the repositioning of the BHs to the nearby potential minimum, the BHs will have well-defined velocities at each time step (this is true whether or not we add the dynamical friction). This allows us to apply further merging criteria based on the velocities and accelerations of the black hole pair, and thus avoid early mergers of gravitationally unbound pairs.

We follow \cite{Bellovary2011} and \cite{Tremmel2017}, and use the criterion
\begin{equation}
    \label{eq:merge_criterion}
    \frac{1}{2}|\bf{\Delta v}|^2 < \bf{\Delta a}\cdot \bf{\Delta r}
\end{equation}
to check whether two black holes are gravitationally bound. Here $\bf{\Delta a}$,$\bf{\Delta v}$ and $\bf{\Delta r}$ denote the relative acceleration, velocity and position of the black hole pair, respectively. Note that this expression is not strictly the total energy of the black hole pair, but an approximation of the kinetic energy and the work needed to get the black holes to merge. Because in the simulations the black hole is constantly interacting with surrounding particles, on the right-hand side we use the overall gravitational acceleration instead of the acceleration purely from the two-body interaction.

\subsection{Dual and Offset AGN Selection}
\label{subsec:criterion}
Among all MBHs in the simulation at a fixed redshift, we define an MBH pair as two MBHs with a separation $\Delta r<30\,{\rm kpc}$ (proper).
For this work, we only focus on the massive end of our population by restricting to MBH pairs with both MBHs above $10^7\,M_\odot$ for two main reasons:
first, with the current dynamical-friction model in \texttt{Astrid}, there remain large uncertainties in the dynamics of lower-mass MBHs due to the introduction of the dynamical mass;
second, the low-mass/faint-end luminosity function from most hydrodynamical simulations are high compared with observations, so that including the $M_{\rm BH}<10^7\,M_\odot$ MBHs may leads to an over-estimation of dual AGN passing certain luminosity thresholds.

Among the large MBHs, we then define AGN as MBHs with bolometric luminosity $L_{\rm bol}>10^{43}\,\mathrm{erg/s}$.
Dual AGN are MBH pairs in which both of the MBHs are AGN. 
We also define a population of offset AGN following the definition in \cite{Steinborn2016} and \citet{DeRosa2019}, where only one of the MBHs in the pair is an AGN.
We note that sometimes in the (observation) literature, "offset AGN" also refers to single AGNs which are offset from the galaxy center.
For the inactive MBHs in the offset pairs, we also require them to have masses $>10^7\,M_\odot$. 
This is to separate the MBHs that only become inactive during the galaxy merger from a much larger population of low-mass companions around the bright AGN.
% \yueying{Why the $>10^7\,M_\odot$ criteria is related to galaxy merger? }

After applying the above criterion, there are 2008 (439) MBH pairs with $M_{\rm BH}>10^7\,M_\odot$ at $z=2$ ($z=3$), among which 1087 (329) are dual AGN, 842 (110) are offset AGN, and 79 (10) are no-AGN pairs.
For the two MBHs in the pair, we will refer to the more massive one \textit{at the time of observation} as the primary MBH (or BH1), and the less massive one as the secondary MBH (or BH2).
Note that we assign the primaries and secondaries by the MBH masses instead of the luminosities.
We will refer to the more luminous MBH as the brighter AGN, and the less luminous MBH as the fainter AGN.
We identify the host galaxies of the MBHs with \texttt{Subfind}, but note that during the close encounters of galaxies, \texttt{Subfind} may not be able to separate the merging systems well. 
This is especially the case for offset AGN hosts since the gas and stellar disruption is very strong.
Finally, when tracing the MBH and galaxy properties back in redshift, we always follow the more massive progenitor if the MBH of interest has gone through prior mergers.

In Figure \ref{fig:img}, we show the dual and offset AGN in a $250\,{\rm cMpc}/h\times 150\,{\rm cMpc}/h \times 20\,{\rm cMpc}/h$ slice of the \texttt{Astrid} simulation. 
The background image shows the gas distribution of the simulation, color-coded by the gas temperature where warmer regions corresponds to higher temperatures. 
At the position of each pair, we zoom into the IGM and galaxies surrounding the AGN pairs.
The yellow dots mark all AGN with $M_{\rm BH}>10^7\,M_\odot$ and $L_{\rm bol}>10^{43}\,\mathrm{erg/s}$.
Note that we have shown all of the dual and offset AGN in this slice, and we can see that the distribution of duals and offsets is a sparse representation of the underlying galaxy/AGN distribution.

\section{Properties of High-z Dual and Offset AGN}
\label{sec:z=3}
In this section, we investigate the static properties of the dual and offset pairs selected at $z=2$ at the time of observation. 
Specifically, we will look at the separation of the pairs, their mass/luminosity function, Eddingon ratio and host galaxy properties compared with the underlying AGN population.

\subsection{Dual Fraction}
\label{sec:dual_frac}

\begin{figure}
\centering
  \includegraphics[width=0.49\textwidth]{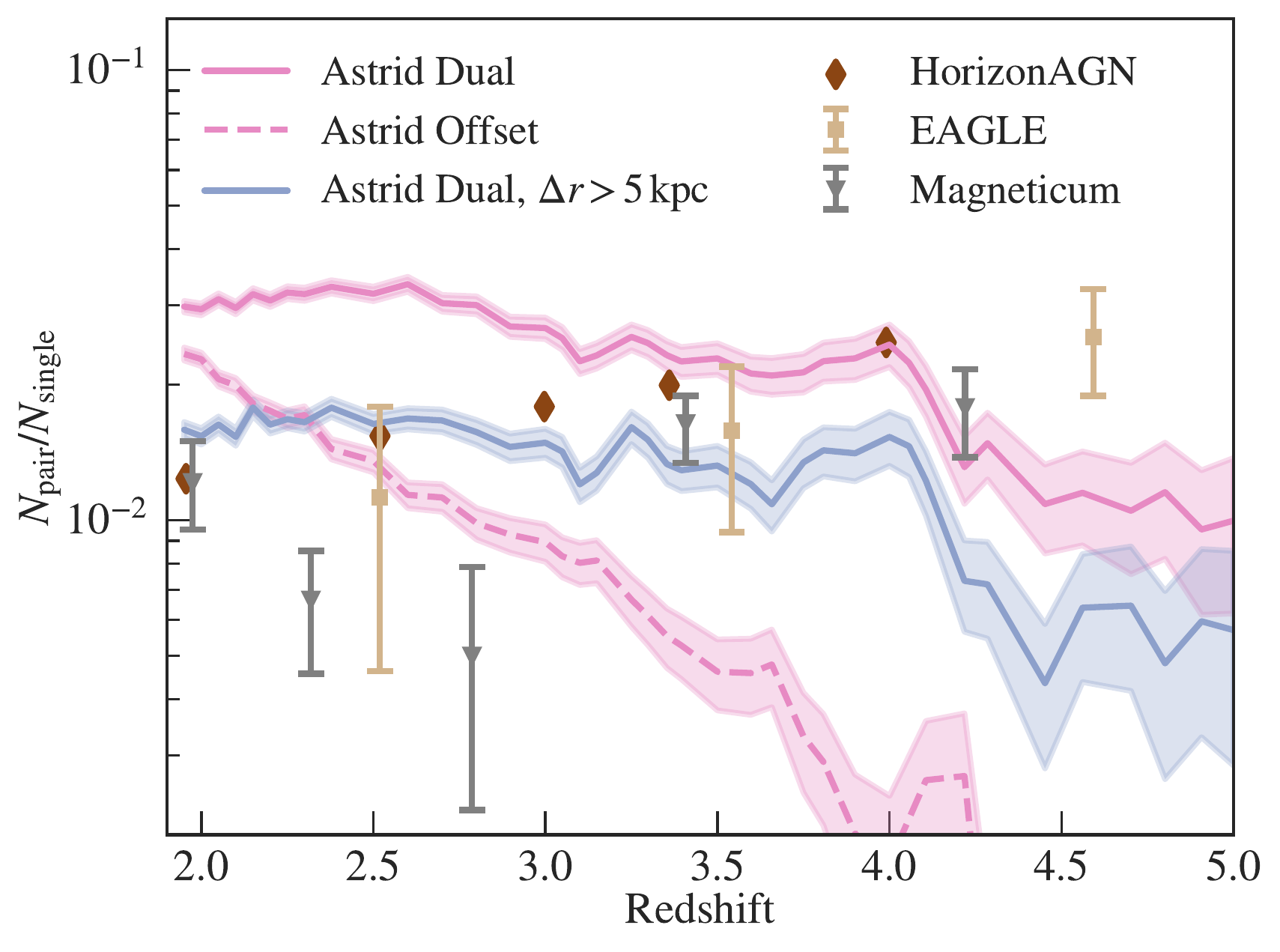}
  \caption{Fraction of dual/offset AGN among the underlying massive AGN population (\textit{\textbf{solid/dashed pink}}). To compute the fraction, we divide the number of dual/offset AGN by the number of massive, luminous MBHs ($M_{\rm BH}>10^7 M_\odot$, $L_{\rm bol}>10^{43}\,\mathrm{erg}/s$). We also show the dual fraction with a selection criterion of  $5\,{\rm kpc}<\Delta r<30\,{\rm kpc}$ ({\textit{\textbf{purple}})}.
  For comparison, we plot the dual fractions in recent simulation works of comparable box sizes including EAGLE (\textit{\textbf{beige square}}), HorizonAGN (\textit{\textbf{brown diamond}}), and Magneticum (\textit{\textbf{grey triangle}}).
  }
  \label{fig:redshift_ev}
\end{figure}

The fraction of dual AGN relative to the underlying single AGN population could be a proxy for the number of massive galaxies under-going galaxy mergers.
Observational studies suggest that the fraction of dual AGN is small \citep[e.g.][]{Fu2011,Rosario2011}.
The dual AGN fraction in the local Universe has been estimated from the dual AGN sample of \citet{Koss2012} detected with X-ray spectroscopy to be about $2\%$.
\citet{Liu2011} used a sample from the  Seventh  Data  Release of the SDSS survey at $z=0.1$ based on diagnostic emission-line ratios and estimated a dual AGN fraction with  $\Delta r<30\,{\rm kpc}$ to be  $1.3\%$.
Constraints on the evolution of dual AGN fraction at higher redshifts is still an ongoing work: recently \citet{Silverman2020} found a dual quasar fraction of $0.26 \pm 0.18\%$ from $z=3$ to $z=1.5$, with no evidence for a redshift evolution \citep[also see][for the fraction of bright double quasars at $z\sim 2$]{Shen2022}. 
However, these high-redshift double quasar candidates are bright compared to our samples, so we do not directly draw comparison between our dual fraction and those computed from these observation works.
Currently, most of the redshift evolution estimates of the dual fraction still come from cosmological simulations.

In Figure \ref{fig:redshift_ev}, we show the redshift dependence of the dual and offset AGN fraction in the \texttt{Astrid} simulation.
To compute the dual (offset) fraction, we take the number of dual (offset) AGN selected based on the criterion in Section \ref{subsec:criterion} as the numerator, and take all AGN with $M_{\rm BH}>10^7\,M_\odot$, and $L_{\rm bol}>10^{43}\,\mathrm{erg}/s$ as the denominator.
From $z=4$ to $z=3$, the dual fraction shows no redshift dependence and remains $\sim 2.3\%$.
After $z=3$, we see a slight rise in the fraction of duals to $\sim 3\%$.
Above $z=4$, there is a drop in the dual fraction to $\sim 1\%$.
The fraction of offset AGN increases significantly from $z=4$ to $z=2$, likely due to the increasing number of minor galaxy mergers which is the major sight for the offsets \citep[see e.g. Section \ref{sec:mass-lum}, also][]{Tremmel2018,Ricarte2021}.

We compare our result to the dual fraction estimates from previous numerical works at high redshifts \citep[e.g.][]{Volonteri2015,Steinborn2016,Rosas-Guevara2019}.
The value of our dual fraction generally agrees with those calculated from other simulations, although at $z\sim 2$ our estimated dual fraction is two times higher.
In contrast to these works which found a decrease in the dual fraction with time, our dual fraction is mostly independent of redshift and increases slightly from $z=3$ to $z=2$.
We note that the data from the simulations shown are not based on the same dual AGN and AGN selection criterion.
For example, the \texttt{EAGLE} sample does not include any pairs with $\Delta r<1\,{\rm kpc}$, and is not subjected to the $M_{\rm BH}>10^7\,M_\odot$ mass cut.
The underlying AGN population is also different in their masses and luminosities, due to the BH model used in each simulation \citep[e.g.][]{Habouzit2022}.

Another difference between the simulations is the spatial resolution and MBH merging criterion, which can affect the number of small-separation pairs. 
Among the simulations shown in Figure \ref{fig:redshift_ev}, \texttt{Astrid} has the most strict MBH merging criterion, and thus we expect more dual AGN with $\sim {\rm kpc}$ separation from \texttt{Astrid}.
This could be one reason for our higher dual AGN fraction.
To show the resolution dependence of the dual fraction, we also show the fraction calculated only using duals separated by $\Delta r>5\,{\rm kpc}$.
With this selection criterion, the dual fraction systematically drops by $\sim 50\%$ at all redshifts. 
The redshift-dependency of our dual fraction is not affected by excluding the close-separation pairs, and compared with other simulations, we produce a larger fraction of dual AGNs at $z<3$.
This can be a result of the velocity-based merging criterion we have adopted, which was absent from most previous simulations.
It has been shown that applying the velocity-based merging criterion can uniquely lead to long lived pairs of MBHs at galaxy-scale separations \citep[e.g.][]{Tremmel2018,Tremmel2018b,Barausse2020,Chen2021}, such that a larger fraction of high-redshift pairs remains observable as dual AGNs at $z\sim 2$.

\begin{figure*}
\centering
  \includegraphics[width=0.48\textwidth]{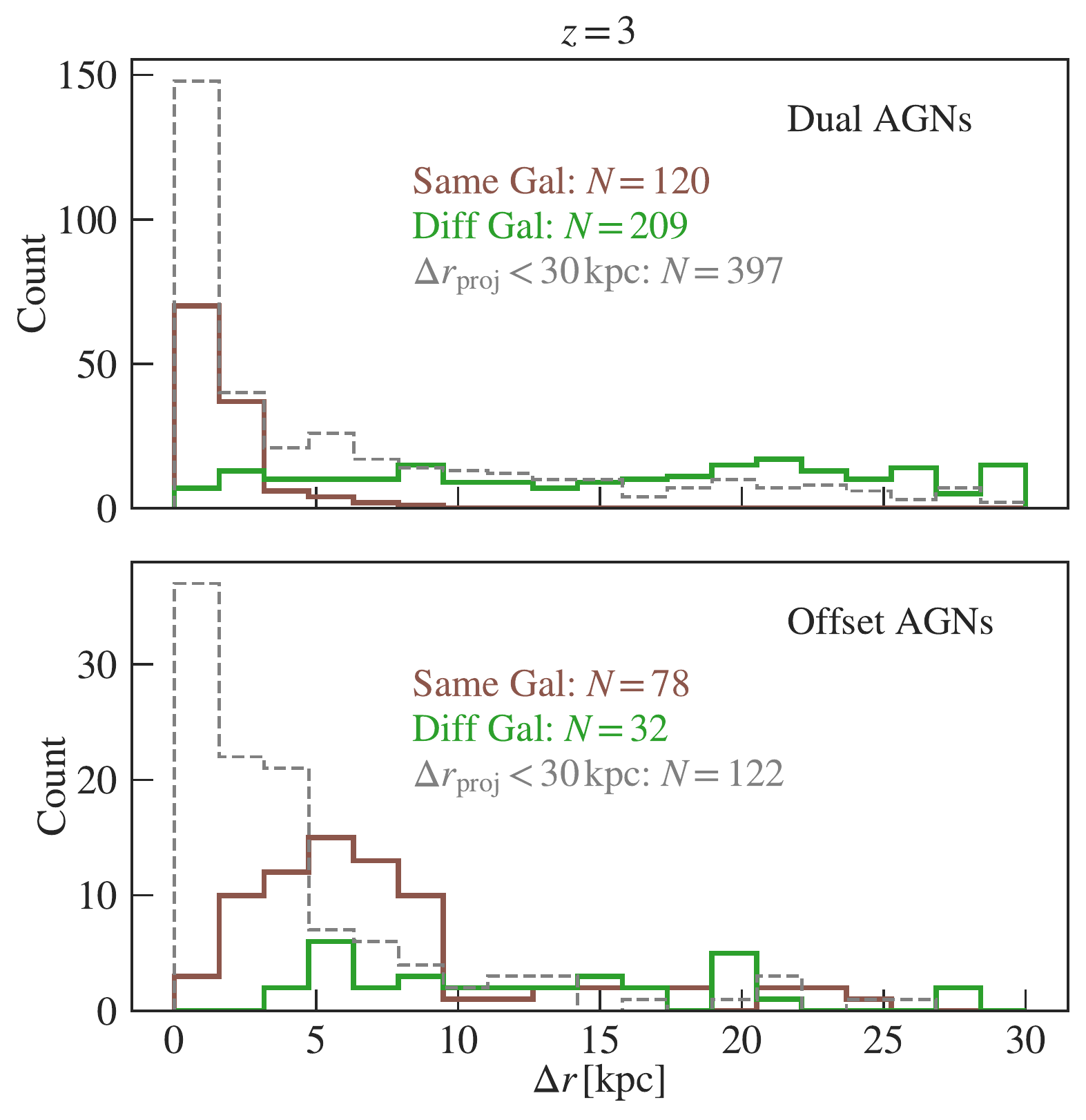}
  \includegraphics[width=0.48\textwidth]{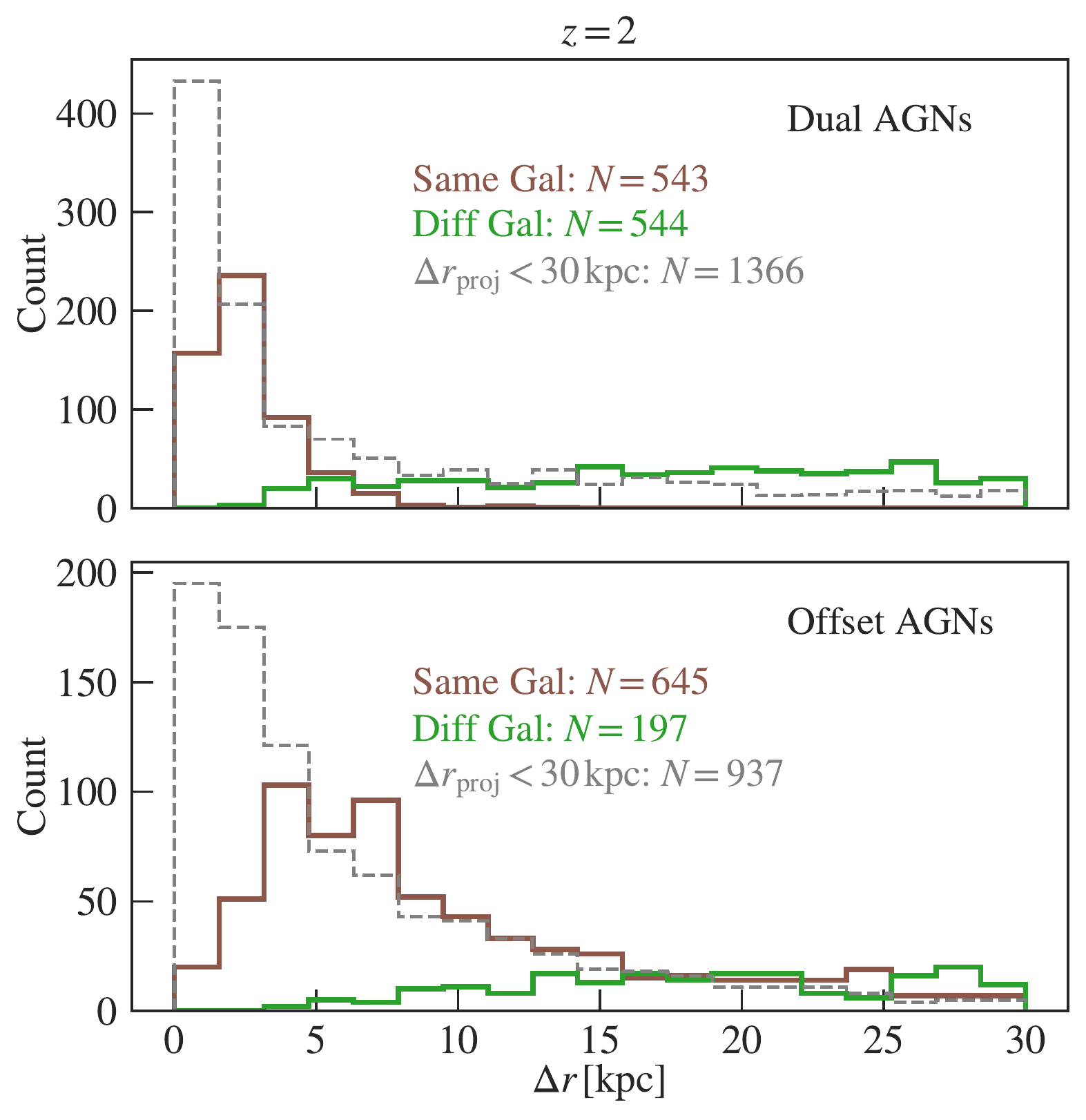}
  \caption{Distribution of the distance between the two MBHs in the dual AGN (\textit{\textbf{top}}) and offset AGN (\textit{\textbf{bottom}}) at $z=3$ (\textit{\textbf{left}}) and $z=2$ (\textit{\textbf{right}}).
  Here we separate each population by whether the two MBHs are embedded in the same galaxy (\textit{\textbf{brown}}) or not (\textit{\textbf{green}}). 
  Between $z=3$ and $z=2$, the number of duals goes through a three-fold increase, whereas the number of offsets becomes seven times larger.
  We also show the dual and offset AGN selected based on the 2D projected distance instead of the true distance, to mimic the selection from observations (\textit{\textbf{grey dashed}}). For those pairs, the x-axis represents the projected distance. 
  }
  \label{fig:distance}
\end{figure*}

\subsection{Pair Separations}

In Figure \ref{fig:distance}, we plot the distributions of the separation between the two MBHs in the dual and offset AGN pairs at $z=3$ and $z=2$.
Among duals and offsets, we further categorize the pairs as same and different galaxy duals (offsets) according to whether the two MBHs are in the same galaxy or not at the time of observation, where the host galaxies are identified with the subhalo catalog generated by \texttt{Subfind}.
Since our simulation adopts a sub-grid dynamical friction model with a stricter merging criterion than most previous works of similar resolutions, we are able to identify more dual and offset AGN at closer separations, down to $\Delta r<1\,{\rm kpc}$.

On the top panels, we show the separation of dual AGN with $\Delta r<30\,{\rm kpc}$.
The dual populations at $z=2$ and $z=3$ share many common features. 
For example, almost all of the same-galaxy dual AGN have separations below $\Delta r=5\,{\rm kpc}$, with a peak near $\Delta r=2\,{\rm kpc}$.
Within our sample, the probability of seeing a $\Delta r<2\,{\rm kpc}$ dual AGN is five times the probability at larger separations.
Finally, the distribution appears flat at $\Delta r>5\,{\rm kpc}$, showing no preferred separation for the different-galaxy duals during the galaxy merger.
In previous works, \cite{Rosas-Guevara2019} find a peak of dual separations within $[20\,{\rm kpc},25\,{\rm kpc}]$, but they did not consider any pairs below $5{\rm kpc}$.
\cite{Steinborn2016} and \cite{Volonteri2021} also uses various models for the sub-grid dynamical friction, and found a higher probability density of duals at $\Delta r< 5 {\rm kpc}$.

On the bottom panels of Figure \ref{fig:distance}, we show the separation distribution of the offset AGN pairs.
Contrary to the dual AGN which accumulates near separation of $1\,{\rm kpc}$, there is almost no offset AGN at such close separations.
Instead, most offset AGN are found at separations around $5\,{\rm kpc}$.
As we will see in the later sections, this is mainly because the stripping of the secondary host galaxy is most severe when the two merging galaxies are separated by around $5\,{\rm kpc}$, causing the secondary AGN to lose its gas supply.
When the separation of the MBHs gets closer to $\sim 1\, {\rm kpc}$, however, the secondary MBH begins to accrete from the gas in the primary galaxy, thereby turning the offset pair into a dual at the $1\,{\rm kpc}$ separation. 

By comparing the $z=2$ and $z=3$ population, we also find interesting differences.
Out of the dual AGN at $z=3$, only $30\%$ reside in the same host galaxy and are within $\Delta r <5\,{\rm kpc}$ in separation. 
However, at $z=2$, $\sim 50\%$ reside in the same galaxy and $\sim 50\%$ in different galaxies.
The evolution from $z=3$ to $z=2$ also saw a large increase in the fraction of offset AGN: while the number of duals have increased by $\sim 200\%$, the offset AGN has grown by $\sim 700\%$.
The growth of the same-galaxy, close duals as well as offsets is a result of the dynamical friction model and merger criterion in \texttt{Astrid} which prevent dual AGNs from merging immediately after the host galaxy merger (also see the detailed discussion in Section \ref{sec:dual_frac}).
During the dynamical-friction dominated orbital decay, an increasing number of AGN go though gas stripping and become offset AGN instead of duals.

Also in Figure \ref{fig:distance}, we show the pairs selected with the 2D projected separation $\Delta r_{\rm proj}$, rather than the true separation, to mimic the selection function of observations.
Here we take $\Delta r_{\rm proj}$ to be the projection of $\Delta r$ on the $x-y$ plane, and we also limit the projection depth to be $|\Delta z|<100\,{\rm kpc}$.
For dual AGN, using the projected separation increases the selected pairs by $\sim 30\%$, meaning that at separations of $\Delta r>30\,{\rm kpc}$, there are still a significant number of pairs residing in separated galaxies.
Furthermore, using projected separation also increases the probability of pairs at $\Delta r_{\rm proj}<15\,{\rm kpc}$.
For the offset pairs, however, $\Delta r_{\rm proj}$-based selection only includes $<10\%$ more pairs, which is a lot less compared with the increase in the dual AGN.
The reason is that with a mass cut of $M_{\rm BH}>10^7 M_\odot$, the secondary MBHs rarely fall below $L_{\rm bol}<10^{43}\,\mathrm{erg}/s$ without severe disruption of the gas during galaxy mergers (see e.g. Figure \ref{fig:mass_lum}).
Hence, at $\Delta r>30\,{\rm kpc}$ and $\Delta r_{\rm proj}<30\,{\rm kpc}$ when the host galaxies have barely interacted, massive but inactive secondaries are hard to find.
Another consequence of the lack of large-separation offset pairs is that using the projected distance will significantly bias the observed distribution towards the lower end:
$\Delta r_{\rm proj}$ has a large peak at $\sim 1\,{\rm kpc}$, although the true separations rarely fall into this bin.

\subsection{Mass and Luminosity}
\label{sec:mass-lum}
\begin{figure*}
\centering
  \includegraphics[width=1\textwidth]{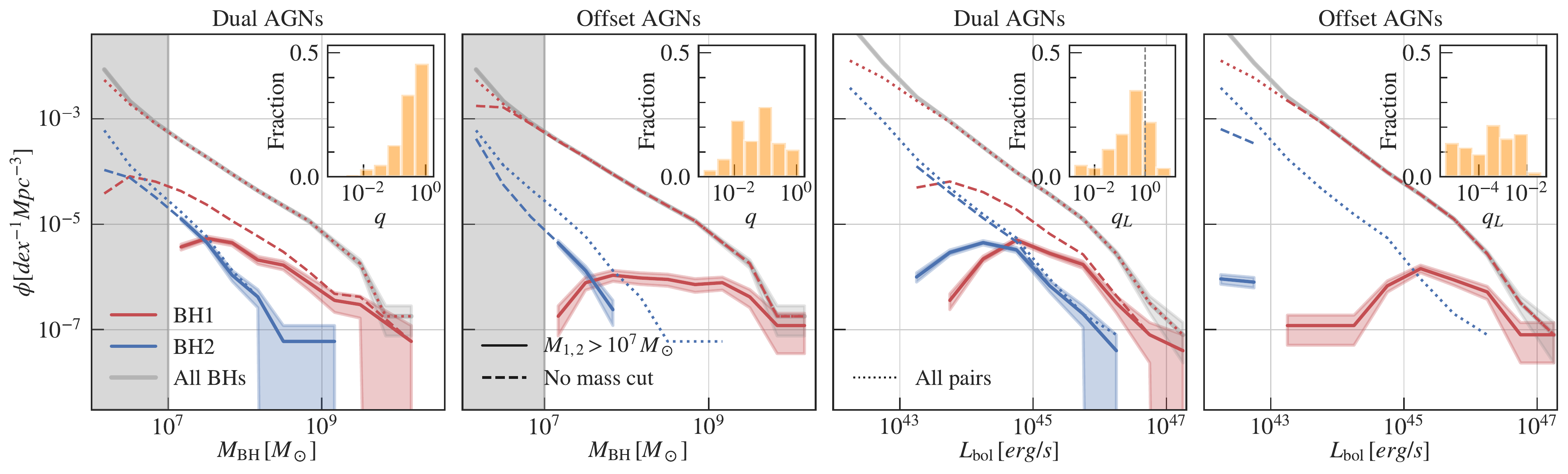}
  \caption{Mass functions (\textit{\textbf{left two columns}}) and luminosity functions (\textit{\textbf{right two columns}}) of the dual and offset AGN, compared with the underlying MBH population and MBH pairs. The \textit{\textbf{solid red (blue)}} shows the distribution of the more (less) massive MBH in the dual/offset AGN pairs. The mass and luminosity function of all MBHs are shown in \textit{\textbf{solid grey}} (we start the bins at $M_{\rm BH}=10^6\,M_\odot$ and $L_{\rm bol}=10^{42}\,\mathrm{erg}/s$ for a clearer view of the high-mass end). To illustrate the effect of our mass threshold at $M_{\rm BH}>10^7\,M_\odot$, we also show the "dual" and "offset" pairs without applying the mass threshold (\textit{\textbf{dashed}}). Note that for such "duals", the luminosity threshold of $L_{\rm bol,1,2}>10^{43}\,\mathrm{erg}/s$ is still present, while for the "offsets", the only constraint is $L_{\rm bol,1}>10^{43}\,\mathrm{erg}/s$. Finally, we show the mass and luminosity functions of all MBH pairs with $\Delta r < 30\,{\rm kpc}$ (\textit{\textbf{dotted}}). 
  The inset panels show the mass and luminosity ratio between the less massive MBHs and the more massive MBHs in the dual and offset pairs.
  Offset AGN have greater mass and luminosity contrasts compared with duals.
  Also note that for duals, the more massive MBHs do not necessarily correspond to the brighter AGN.}
  \label{fig:mass_func}
\end{figure*}

Figure \ref{fig:mass_func} shows the mass and luminosity functions of the two MBHs involved in dual and offset AGN pairs at $z=2$, in comparison with the underlying single MBH population and the MBH pairs without the mass/luminosity cuts.
By comparing the thin dotted lines which include all MBH pairs with the solid brown line showing the single MBH distribution, we can see that almost all MBHs with $M_{\rm BH}>10^6\,M_\odot$ or $L_{\rm bol}>10^{43}\,\mathrm{erg}/s$ have a companion black hole, typically with a much smaller mass.
We note that to avoid double counting MBHs involved in multiplets when calculating the mass and luminosity functions, we only count each MBH once by taking unique IDs. 
For this work, we do not explicitly search for MBH multiplets, so our catalog is subject to double-counting in the case of multiplets (this affects $\sim 10\%$ of all the $M_{\rm BH}>10^7\,M_\odot$ pairs), but for BH statistics, we always take unique IDs in case one BH is involved in more than one pair.

After applying a luminosity threshold of $L_{\rm bol}>10^{43}\,\mathrm{erg}/s$ to both MBHs in the "dual" AGN case, we see a significant drop in the pair fraction shown by the dashed lines.
This implies that even though all massive MBHs have a close companion, only $\sim 10\%$ have a companion that is also luminous.
For the "offset" AGN, however, since we only apply the luminosity threshold to the more luminous MBH in the pair, the mass and luminosity distributions of the primary MBH are almost not affected at the high-mass end.
For the less-luminous MBH in the "offset" pair,  selecting only the luminous primary suppresses the mass and luminosity function of the secondary.
Requiring the secondary to be under-luminous also suppresses the high-mass end, as these pairs would fall into the "dual" AGN category.
Note that since this is not how we define our dual and offset AGN for this work, we have added quotes when referring to "duals" and "offsets" selected only by their luminosities but not their masses.

Finally, we show the mass and luminosity functions of our dual and offset catalog in solid lines.
When adding the $M_{\rm BH}>10^7\,{M_\odot}$ threshold to both MBHs in the pair, the dual fraction dropped to $\sim 3\%$.
The change in the mass and luminosity functions are not greatly affected by the additional cut on the mass, since among MBHs with luminosities above $10^{43}\,\mathrm{erg}/s$, most already have $M_{\rm BH}>10^7\,M_\odot$.
The effect on the offset pairs is more significant. Requiring the less luminous MBH in the offsets to have $M_{\rm BH}>10^7\,M_\odot$, we are explicitly selecting out the rare population of massive but very under-luminous secondaries.
By making this selection, we can separate out the naturally low-luminosity secondaries due to their low mass, from the under-luminous secondaries that only become faint due to the galaxy merger events.

The mass and luminosity ratios between the two MBHs in the dual and offset pairs are shown in the inset panel of each figure.
Here we take the ratio of both the mass and the luminosity between the less massive and more massive MBHs (in particular, we do not take the luminosity ratio between the fainter and the brighter AGN).
$>80\%$ duals have $q>0.1$, while only $\sim 20\%$ offsets have $q>0.1$.
The luminosity contrast between the two MBHs in the offsets is even greater: the primary AGN are $>100$ times brighter than the secondary among all offsets.
For dual AGN, this feature has already been found in observational works such as \citet{Koss2012}, and been seen in various simulations \citep[e.g.][]{Callegari2009,Steinborn2016,Capelo2017}.

% Comparing the mass functions of the dual and offset pairs, we see a clear feature that offset AGN have larger mass ratios.
% In our work, since we have a much larger sample of dual and offset AGN, we can see more clearly the statistical difference in mass ratios: the primary MBHs in offsets shows a very flat mass distribution, with a drop below $10^8\,M_\odot$, while the secondary MBHs with masses all below $10^8\,{M_\odot}$.
% For the dual AGN, the primaries and secondaries have similar mass distributions.

%----------------------------------------------------------------

\begin{figure}
\centering
  \includegraphics[width=0.5\textwidth]{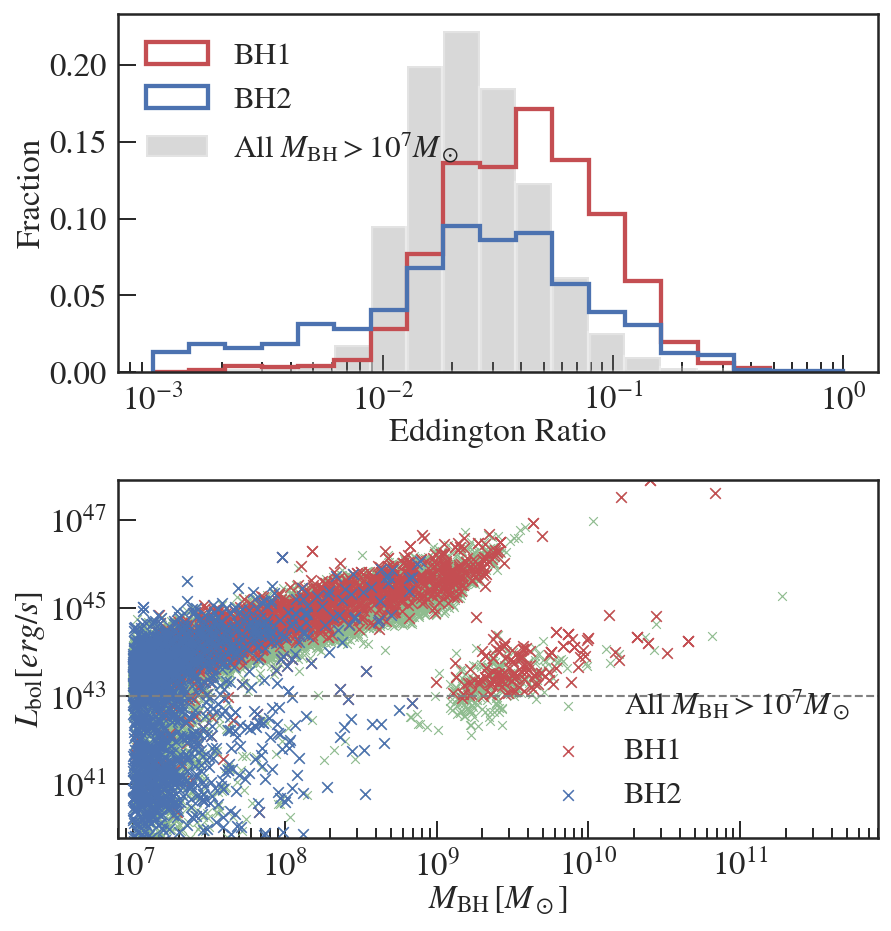}
  \caption{
  \textit{\textbf{Top:}} Eddington ratio of the dual and offset pairs (with the more massive one shown in \textit{\textbf{red}}, and the less massive one shown in \textit{\textbf{blue}}), compared with the underlying massive MBHs with $M_{\rm BH}>10^7M_\odot$ (\textit{\textbf{grey}}, adding the extra $L_{\rm bol}>10^{43}\,\mathrm{erg/s}$ constraint does not affect the peak of the distribution).
  \textit{\textbf{Bottom:}} Masses and luminosities of the MBHs in pairs, plotted on top of all MBHs with masses above $10^7\,M_\odot$ (\textit{\textbf{green}}).
  The horizontal dashed line marks the threshold for an AGN, and the points below it are the secondary MBHs in an offset AGN pair.
  We can see that almost all the $L_{\rm bol}<10^{43}\,\mathrm{erg/s}$ MBHs with $M_{\rm BH}>10^7\,M_\odot$ are involved in an offset pair.
  }
  \label{fig:mass_lum}
\end{figure}

\begin{figure}
\centering
  \includegraphics[width=0.5\textwidth]{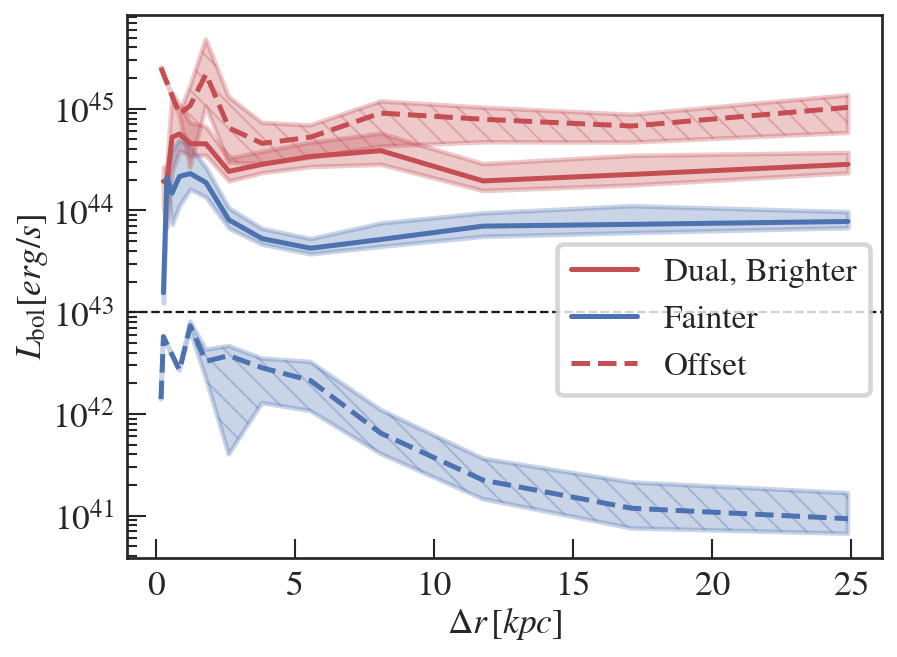}
  \caption{The relation between the pair separation and MBH luminosities. At large separations ($\Delta r > 10\,{\rm kpc}$), the luminosities are not sensitive to the separation. For closer pairs, the luminosity of the fainter MBH is inversely correlated with $\Delta r$.
  }
  \label{fig:dr_lum}
\end{figure}

\begin{figure*}
\hbox{\hspace{-1.5cm} 
  \includegraphics[width=1.2\textwidth]{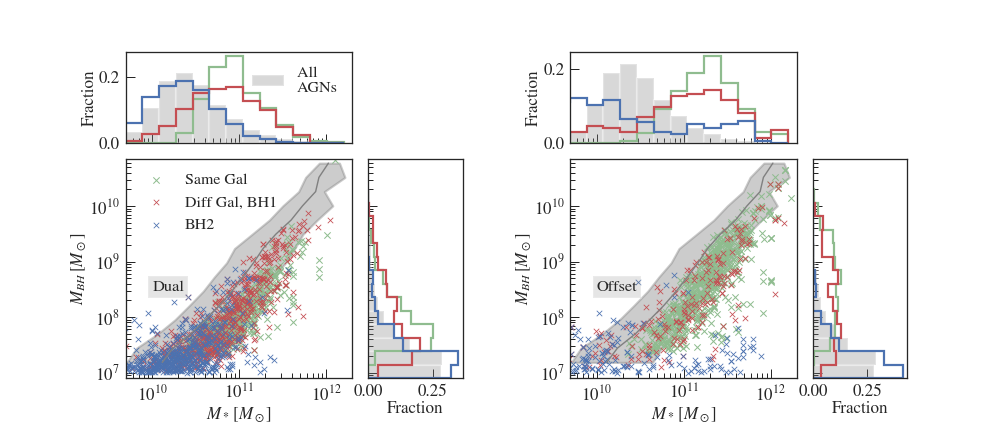}}
  \caption{The $M_{\rm BH}-M_*$ relation of dual AGN (\textit{\textbf{left panels}}) and offset AGN (\textit{\textbf{right panels}}). For different-galaxy pairs, we show the more massive MBH in each pair in \textit{red} and the less massive one in \textit{blue}. For same-galaxy pairs, we plot the sum of the MBH masses against their host galaxy mass in \textit{green}. The \textit{grey} line shows the median BH mass of all MBHs with $M_{\rm BH}>10^7\,{M_\odot}$ The side panels shows the 1D distribution of the MBH masses and galaxy masses.}
  \label{fig:mstar}
\end{figure*}

The top panel of Figure \ref{fig:mass_lum} shows the distribution of the Eddington ratios of the pairs at $z=2$ compared with all MBHs with $M_{\rm BH}>10^7\,M_\odot$ (note that adding an extra $L_{\rm bol}>10^{43}\,\mathrm{erg/s}$ threshold to the underlying single AGN population does not change our conclusion here, so we do not show an extra line for that population).
We can see that compared to the overall AGN population at the same masses with a typical Eddington ratio of $\sim 0.025$, the pairs have a higher level of activation, where the Eddington ratios peak above $0.05$.
On the bottom panel,  we show the mass-luminosity relation for the dual and offset pairs, plotted on top of all MBHs.
The primary MBHs follow the underlying MBH distribution but appear slightly over-luminous compared to the mean relation of the non-pair population.
The blue crosses mark the secondary MBHs in the pair, and the ones falling below the dashed line are the secondaries of offset AGN.
Comparing these secondaries to the overall MBH population, we can see that the inactive MBHs of the offsets are extremely under-luminous.
In fact, the offset secondaries appear to be the majority of the $L_{\rm bol}<10^{43}\,\mathrm{erg/s}$ MBHs with $M_{\rm BH}>10^7\,M_\odot$.
It is very rare for an $M_{\rm BH}>10^7\,{M_\odot}$ MBH to have $L_{\rm bol}<10^{43}\,\mathrm{erg}/s$ if it is not involved in a galaxy merger.
Finally, we note that another group of low-Eddington ratio MBHs are the heaviest MBHs with $M_{\rm BH}\gtrsim 10^9\,M_\odot$.
This is due to the kinetic AGN feedback that actively suppresses the gas accretion among the most massive BHs.

The enhancement in the AGN activation among pairs shown above is usually attributed to the gas-inflow during galaxy mergers. 
Previous simulation works have seen peaks in the pair activation at $1\sim 10\,{\rm kpc}$ \citep[][]{Wassenhove2012} and $0.1\sim 2\,{\rm kpc}$ \citep[][]{Blecha2013}.
Recent observations by \citet{Stemo2021} has seen a bump in AGN activation at a bulge separation of $\sim 10\sim 15\,{\rm kpc}$.
Also in the nearby universe, enhanced AGN activation is seen in close pairs with less than tens of kpc separations \citep[e.g.][]{Ellison2011,Liu2012}.

In Figure \ref{fig:dr_lum}, we show the relationship between the offset and dual AGN luminosity and the pair separations.
Here we log-binned the pairs by their separation, and for each bin we plot the median luminosity enclosed by the interval including $80\%$ of MBHs in that bin.
The red (blue) lines correspond to the brighter (fainter) MBH, and the solid (dashed) curves represent the dual (offset) AGN pairs.
For the dual AGN, both the brighter and the fainter AGN exhibit an increase in luminosity at separations below $\sim 4\,{\rm kpc}$.
The median luminosities then drop slightly at a separation around $5\,{\rm kpc}$, but then increase again at around $10\,{\rm kpc}$.
Because of the rise in AGN luminosity at $\Delta r<5{\rm kpc}$, observations targeted at bright quasar pairs could see a larger fraction of close pairs than observations that also includes fainter AGN pairs.
In \citet{Shen2022}, for example, we can see that using a higher luminosity threshold for the dual selection from our simulation than our sample here results in a larger contrast between the number of small-separation duals and large-separation duals.

For the offset pairs, the brighter MBH are more luminous than the duals, and the luminosity depends less on the pair separation.
The fainter MBHs in the offsets, however, show a significant increase in luminosity with smaller separations, and as a result, there are only $<20$ inactive secondaries at a separation below $2.5\,{\rm kpc}$. Our $\Delta r-L_{\rm bol}$ relation suggests that when observations only limit to $\Delta r>5\,{\rm kpc}$ pairs, it may be hard to establish a relation between the AGN activation and the pair separation.

\subsection{Host Galaxies}

% \begin{figure}
% \centering
%   \includegraphics[width=0.45\textwidth]{plots/mass_distribution.pdf}
%   \caption{Comparisons between the galaxy mass (\textit{\textbf{left}})  and the black hole mass (\textit{\textbf{right}}) of the dual (\textit{\textbf{top}}) and offset (\textit{\textbf{bottom}}) AGN with those of all AGN with $L_{\rm bol}>10^{45}\,\mathrm{erg/s}$. }
%   \label{fig:mstar2}
% \end{figure}

\begin{figure}
\centering
  \includegraphics[width=0.5\textwidth]{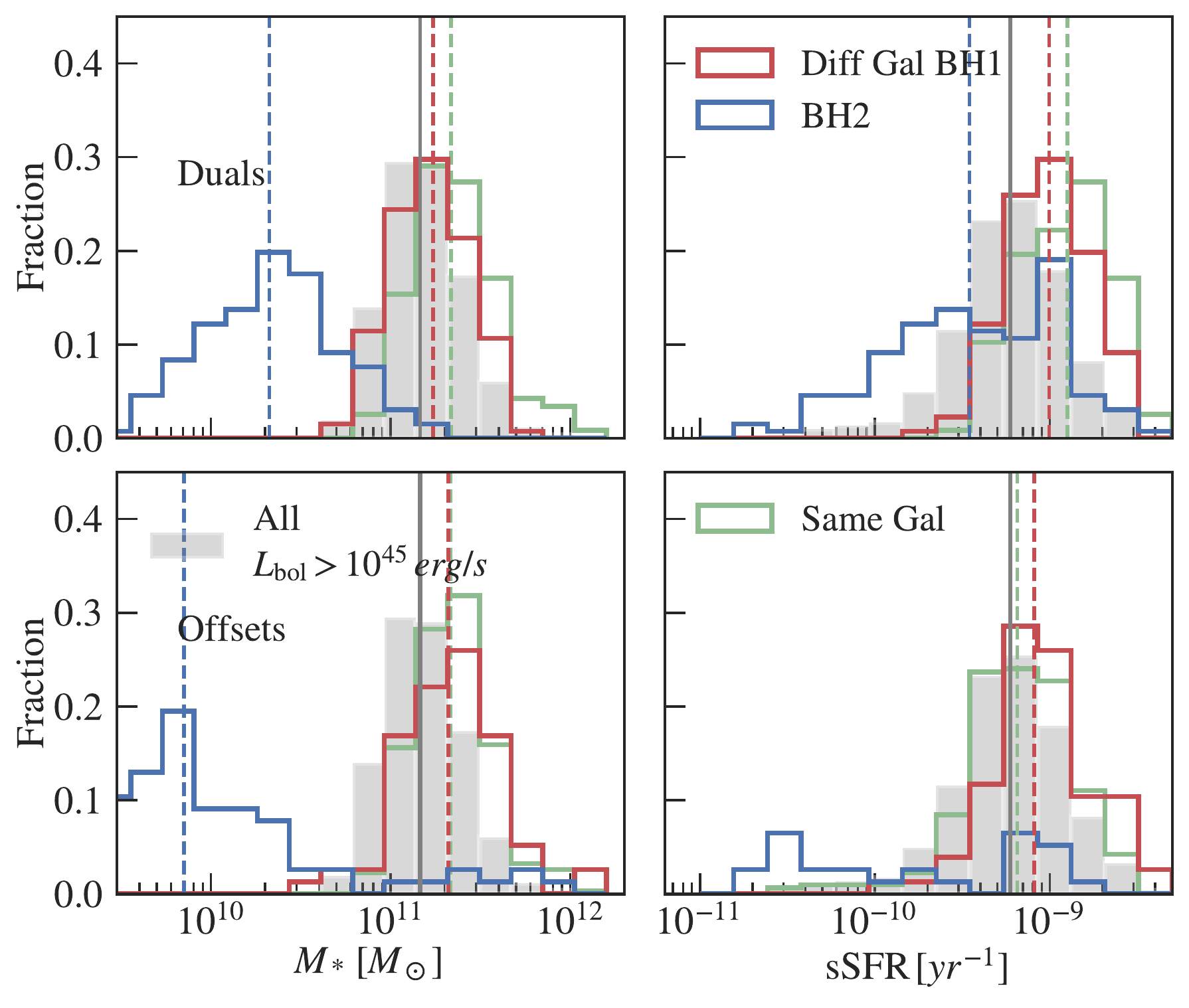}
  \caption{Comparisons between the galaxy mass (\textit{\textbf{left}})  and specific star formation rate (\textit{\textbf{right}}) of the dual (\textit{\textbf{top}}) and offset (\textit{\textbf{bottom}}) AGN with those of all AGN with $L_{\rm bol}>10^{45}\,\mathrm{erg}/s$. For the dual and offset AGNs we have added the $L_{\rm bol}>10^{45}\,\mathrm{erg}/s$ luminosity threshold to the bright AGN in the pairs. The horizontal lines mark the median of each distribution.}
  \label{fig:sfr}
\end{figure}

Figure \ref{fig:mstar} shows the $M_{\rm BH}-M_{*}$ relation for the dual and offset pairs and their host galaxies. 
For the pairs identified in the same galaxy, we plot the total mass of the MBH pair and the mass of their host, while for the different-galaxy pairs, we show the mass of each MBH and host separately.
The central panels show the scattered relation between $M_{\rm BH}$ and $M_{*}$, while the top and right panels show the 1D distributions of the galaxy and MBH masses, respectively.
To compare the duals and offsets with the single AGN population, we show the distribution of all MBHs with $M_{\rm BH}>10^7\,M_\odot$ in grey. 
For the middle panels, the solid curve shows the median galaxy mass within each $M_{\rm BH}$ bin, and the shaded region encloses the scatter of the middle $95\%$ of the galaxy masses in that $M_{\rm BH}$ bin.

From the 1D distributions of the MBH and galaxy masses, we can see that the pairs favor the more massive MBHs and galaxies among the overall MBH population.
For all MBHs selected with $M_{\rm BH}>10^7\,M_\odot$, the galaxy masses center around $2\times 10^{10}M_\odot$. 
For the dual AGN, the host galaxy masses peak near $10^{11}\,M_\odot$, regardless of whether the two MBHs are in the same galaxy.
Part of the reason for the skewed galaxy mass distribution of duals is that the dual AGN selects out primaries that are on the massive end of the single AGN population.
However, when we compare the $M_{\rm BH}-M_*$ relation of duals with that of all massive MBHs, we still see that almost all pairs have galaxy masses above the median relation in the same MBH mass bin, meaning that MBHs in duals are under-massive relative to their host galaxies.
A previous study by \citet{Steinborn2016} of the $z=2$ pairs using the Magneticum simulation has found a similar trend, where their MBH in pairs are systematically under-massive with respect to the host galaxy masses.

For offset AGN, the contrast of the host galaxy masses with those of the single AGN population is even greater.
Most offset AGN reside in galaxies of $M_*\sim 2\times 10^{11}\,M_\odot$.
This is again a consequence of two factors: 
the MBHs with a massive but under-luminous companion are among the most massive BHs, so they naturally reside in large galaxies; galaxy mergers also play a role, because even when compared with other AGN in the same $M_{\rm BH}$ mass bin, the host galaxies of offsets are still larger than the median.
A significant fraction even falls into the top $2.5\%$ of the galaxy mass when compared with their similar-mass single-AGN counterparts.

We notice that there is a contrast between our comparisons of the single-AGN and AGN-pair population and the observational results shown in \cite{Stemo2021}, who found that the MBH mass and galaxy mass distributions of the AGN pairs are not significantly different from the single AGN samples.
One reason for the differences is that our underlying single-AGN samples have a lower $M_{\rm BH}$ distribution compared with their selection function, especially at high redshifts.
In order to mitigate the difference in the underlying AGN sample, in Figure \ref{fig:sfr} we raise luminosity threshold from $10^{43}\,\mathrm{erg}/s$ to $10^{45}\,\mathrm{erg}/s$.
We apply the same luminosity threshold to our dual and offset AGN samples.
The resulting galaxy-mass distribution of the single AGN is closer to the single AGN from \cite{Stemo2021}, which peaks around $10^{11}\,M_\odot$.
After the stricter luminosity cut, we find that the galaxy mass distribution of the duals is similar to that of the single AGN, with the primary galaxy mass slightly higher.
The offset AGN, however, still tend to reside in the high-mass galaxies compared to the underlying single AGN population.

In the right panels of Figure \ref{fig:sfr}, we show the specific star-formation rate (sSFR) for the pairs with $L_{\rm bol,1}>10^{45}\,\mathrm{erg}/s$, compared with all AGN with $L_{\rm bol}>10^{45}\,\mathrm{erg}/s$.
The sSFR is calculated by summing the gas star formation rate within the half-mass radius of the host galaxy, and then dividing it by the total stellar mass within the half-mass radius. 
For the $L_{\rm bol}>10^{45}\,\mathrm{erg}/s$ AGN sample, the sSFR peaks around $0.6\times 10^{-9}\,yr^{-1}$.
The sSFR of the different-galaxy duals is similar to the AGN at similar luminosities and stellar masses, with the primary AGN's sSFR slightly higher.
The hosts of the same-galaxy duals have an overall higher sSFR, with a peak around $10^{-9}\,yr^{-1}$.
Our statistics are consistent with previous studies using idealized galaxy merger simulations \citep[e.g.][]{Wassenhove2012}, who saw peaks in the host galaxies' SFR after a few pericentric passages, when the duals are separated by a few kpcs.
In our case, such duals mostly fall into the same-galaxy dual category.
For the offset pairs, even though the galaxy mass is generally higher compared to the overall luminous AGN, the sSFR of the host galaxies does not differ from the underlying AGN sample.

Comparing our sSFR with the recent observations from \citet{Stemo2021} with similar galaxy masses, we see that our sSFR peaks at a higher value.
Moreover, \citet{Stemo2021} does not see an enhancement in the SFR among the pairs compared to their underlying AGN sample, whereas we see a shift towards higher SFR in our duals.
One reason is that the sSFR increase is most significant for the same-galaxy duals, typically with separations of $\Delta r<5\,{\rm kpc}$.
However, the sample selected based on distinct galaxy bulges from \citet{Stemo2021} consists only of the different-galaxy pairs, among which the increase in star formation has not taken place.

\subsection{Host Halo}

\begin{figure}
\centering
  \includegraphics[width=0.43\textwidth]{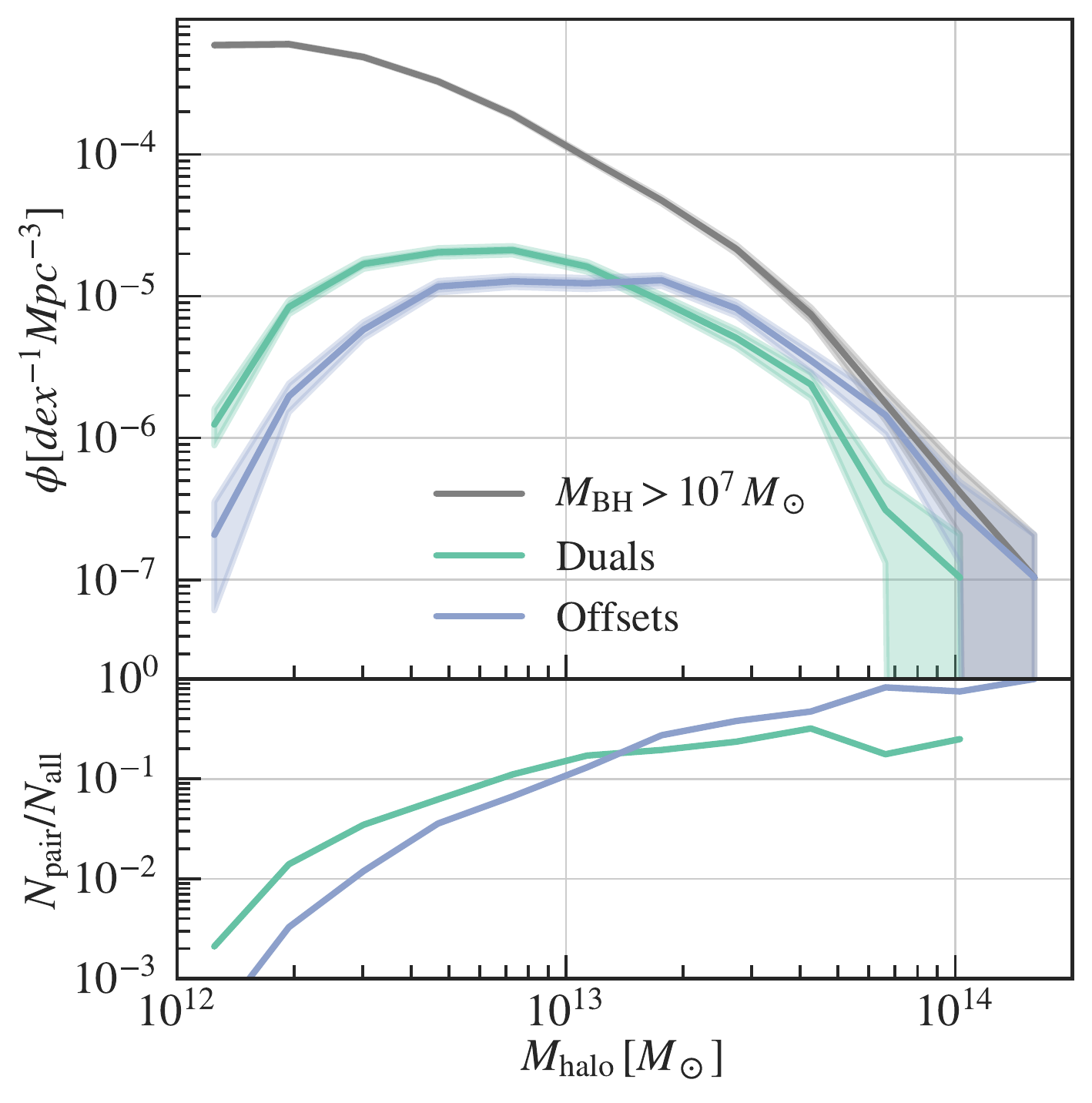}
  \caption{\textbf{\textit{Top:}} Host halo mass functions of the same-galaxy dual (\textit{\textbf{green}}) and offset (\textit{\textbf{purple}}) AGN, plotted with the mass function of all halos hosting at least one MBH with $M_{\rm BH}>10^7\,M_\odot$ (\textit{\textbf{grey}}, and we checked that adding the extra $L_{\rm bol}>10^{43}\,\mathrm{erg/s}$ requirement results in a similar line).
  For hosts of duals and offsets, we only count unique halos, but the fraction of halos hosting two pairs is less than $
  5\%$. 
  \textit{\textbf{Bottom:}} the ratio between the number of dual(offset) host halos and $M_{\rm BH}>10^7\,M_\odot$ MBH host halos in each mass bin.
  }
  \label{fig:mfof}
\end{figure}

Figure \ref{fig:mfof} shows the host FOF halo mass of the dual and offset AGN at $z=2$, together with the host halo mass of all MBHs with $M>10^7\,M_\odot$.
The fraction of dual and offset AGN in each halo mass bin is shown on the bottom panel.
Dual AGN  prefer halos with masses ranging between $10^{12.3}\,M_\odot-10^{13}\,M_\odot$, and are very rarely found in halos with $M_{\rm halo}>10^{13.5}\,M_\odot$.
The offset AGN typically reside in more massive halos compared with the duals, with the majority of them found in halos in the mass range of 
$10^{13}\,M_\odot-10^{14}\,M_\odot$.
The offset fraction increases significantly with the mass of the host halo: $<0.1\%$ of the $\sim 10^{12.2}\,M_\odot$ halos host an offset AGN, whereas $\sim 40\%$ of the most massive halos with $M_{\rm BH}>10^{13.5}\,M_\odot$ host an offset AGN.

One explanation for why we find more offsets than duals in the most massive halos is that the deep potential of such massive halos causes the most gas and stellar disruption of the secondary \citep[also see e.g.][]{Ricarte2021}. 
Therefore, even though the secondary MBH in the offset is initially more massive (as we will show in Section \ref{subsec:pair_evol} and in Figure \ref{fig:mstar_evolve}), it falls victim to the gravitational potential around the primary AGN and ends up lurking in the most massive halos for an extended period of time.

The fact that the most massive halos preferentially host offsets instead of duals also has observational implications:
one way to search for dual AGN is by looking for companion AGN around a sample of single AGN.
At high redshift, the luminosity threshold for detecting AGN is typically high, and thus the resulting observed AGN sample could be embedded in the most massive halos.
These most luminous AGN are more likely to be involved in offset pairs rather than duals, such that its companion may not be detectable through EM observations albeit its high mass.

\subsection{Obscuration of high-redshift pairs}
\begin{figure}
\centering
  \includegraphics[width=0.49\textwidth]{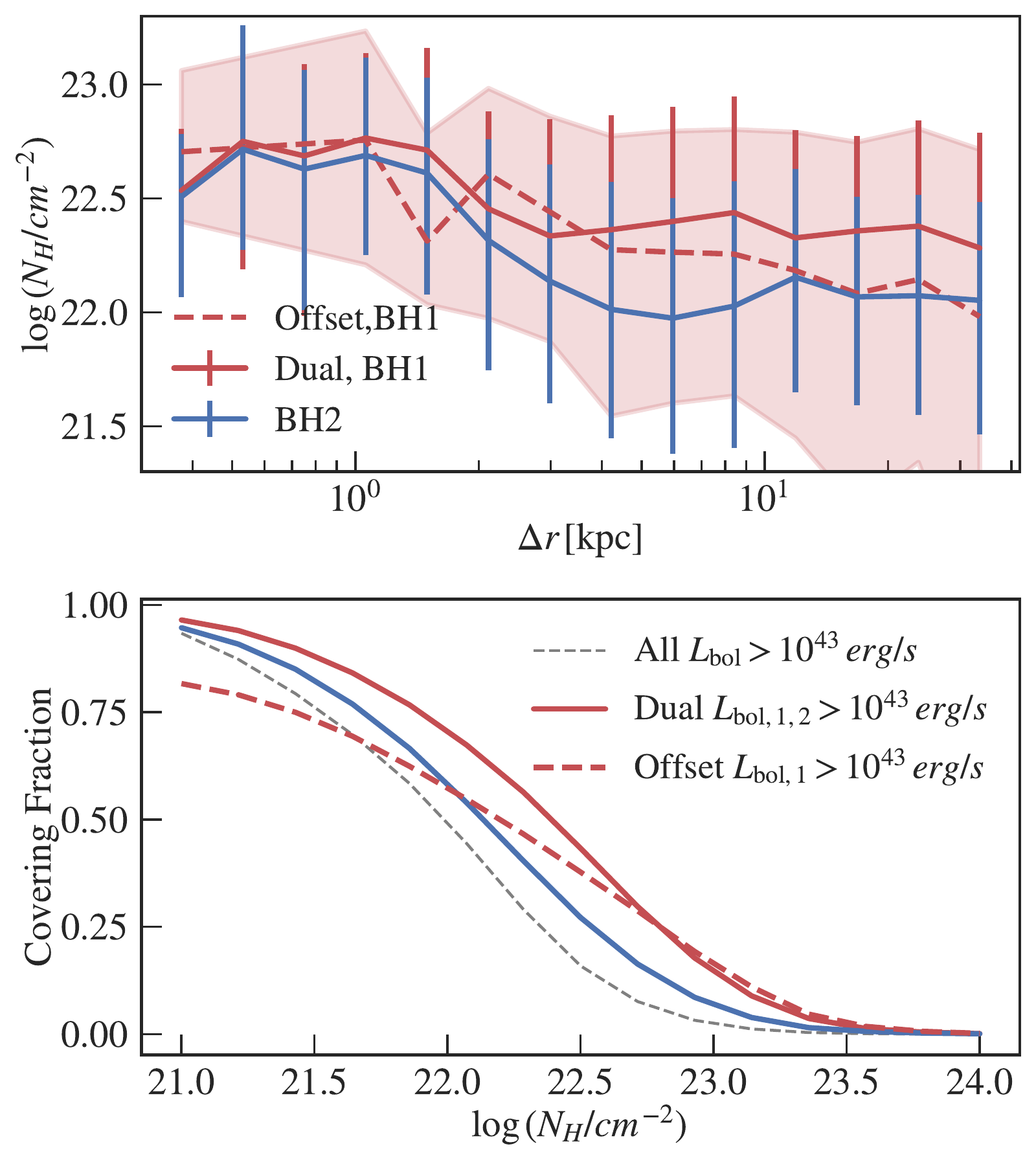}
  \caption{\textit{\textbf{Top:}} The relation between gas column density $N_H$ and pair separations for the dual AGN (the more luminous AGN in \textit{\textbf{red solid}}, the less luminous AGN in \textit{\textbf{blue solid}}) and offset AGN (we only show the active AGN in \textit{\textbf{red dashed}}). For each AGN we compute $N_H$ along 48 random sight lines. The curves show the median $N_H$ of all lines-of-sight in each $\Delta r$ bin, with the shaded area/vertical lines covering the 16 - 84th percentile of the distribution. \textit{\textbf{Bottom:}} The AGN covering fraction of dual and offset AGN assuming various $N_H$ thresholds. To compare with the underlying AGN sample (\textit{\textbf{grey}}), we apply the same $L_{\rm bol}$ lower limit to the pairs. The covering fraction of pairs is higher than the underlying AGN population.}
  \label{fig:obscuration}
\end{figure}
Previous theoretical works have found that the offset in AGN triggering time and the dynamics of the merger are the main factors that explain the paucity of observed AGN pairs \citep[e.g.][]{Wassenhove2012,Blecha2013}.
Besides these factors, the AGN obscuration could also hinder the discovery of dual AGN.
As was discussed in the recent review paper by \citet{DeRosa2019}, some AGN in many confirmed dual systems show no (or very weak) explicit AGN evidence in their optical/near-infrared spectra, indicating that these AGN pairs are heavily dust-enshrouded, and there is also strong evidence that they are also heavily obscured in the X-rays.
It is also known that $20-40$\% of AGN are hidden behind Compton-thick column densities (with $N_{\rm H} > 10^{24}$ $\mathrm{cm}^{-2}$) and $\sim 75$\% of the remaining population are obscured, with $N_{\rm H} = 10^{22} \sim 10^{24} \mathrm{cm}^{-2}$ \citep[e.g.][]{Ueda2014,Buchner2015,Aird2015} at the peak of AGN activity at redshift $z=0.5-3$.
Meanwhile, \citet{Capelo2017} examined the effect of obscuration in hard X-ray luminosities on resolvable scales ($>100\,{\rm pc}$) in their idealized galaxy merger simulations, and found it to be negligible for $z=3$ galaxies.
In this section, we investigate the role of AGN obscuration in pair detection from our sample of dual and offset AGN.

To calculate the column density for each AGN, we follow the method in \cite{Ni2020}, and estimate the contribution to the obscuration only due to the gas in the host galaxy. 
By doing so, we do not account for the AGN obscuration associated with the nuclear torus, on scales of $\sim 10$ pc surrounding the accretion disk of the BH, which is beyond the resolution of cosmological simulations.

In Figure \ref{fig:obscuration}, we show the gas column density of dual and offset AGN binned by the pair separation.
For each AGN we compute $N_{\rm H}$ along 48 random sight lines, and for each $\Delta r$ bin we show the median and the middle $68\%$ of all lines-of-sight within that bin.
For the offset pair, we only show the column density of the primary AGN, as the inactive MBH is likely not observable through EM signatures.
We find that for both dual and offset pairs, the column density increases with decreasing pair separations.
This is particularly true for dual AGN at separations below $\sim 2\,{\rm kpc}$: at this separation, most duals have $N_{\rm H}>10^{22.6}\,{\rm cm}^{-2}$.
Our finding is in line with the recent observational studies by e.g. \citet{Ricci2017}, who find that AGN obscuration reaches its maximum at the late galaxy merger stage, when the nuclei of the two merging galaxies are at a projected distance of $<10\,{\rm kpc}$.
This finding, when combined with the $\Delta r-L_{\rm bol}$ relation in Figure \ref{fig:dr_lum}, shows that the close separation pairs in the post-merger galaxies are both more luminous and more obscured, potentially adding complications the detection of those pairs.

Between the two AGN in a dual, the secondary has a larger increase in the obscuration with decreasing pair separation, as it enters into the gas reservoir of the primary AGN.
Moreover, we also find that the $N_{\rm H}$ of the secondary varies with the angle between the line-of-sight and the dual separation: the $N_{\rm H}$ integrated from the line-of-sight passing near the primary AGN can be three times higher than the $N_{\rm H}$ calculated from the opposite direction, or perpendicular to the dual separation.
One implication is that pairs projected at a smaller separation can have a more obscured secondary, compared to pairs viewed at their true separation.

The bottom panel of Figure \ref{fig:obscuration} shows the AGN covering fraction at different $N_{\rm H}$ thresholds.
Here we take the median overall line-of-sights for all the pairs to compute the covering fraction at each threshold.
To do a fair comparison with the ${\rm log}(L_{\rm bol})>43$ AGN sample, we also apply a ${\rm log}(L_{\rm bol})>43$ lower limit to both AGN of the duals and to the brighter AGN of the offsets.
The AGN covering fraction for the duals is generally higher than that of the underlying AGN population, especially at higher $N_{\rm H}$ thresholds.
At ${\rm log}(N_{\rm H})>23$, only $<3\%$ of the sight lines among all AGN are covered, for dual/offset AGN the fraction ranges from $10\%$ to $20\%$.

% At a first glance, this may seem contradictory to the result from the upper panel, since at the low-luminosity end the column-density of pairs is lower, whereas the covering fraction of the pairs is higher.
% This is mainly because we do not bin by luminosity when calculating the covering fraction, and pairs, in general, occupy the high-luminosity end of the distribution compared to the underlying AGN population (see Figure \ref{fig:mass_func}).
% Hence, at a threshold of e.g. $N_H=10^{23}\,{cm^{-2}}$, although the fraction of covered sight lines in the $L_X={45}\,\mathrm{erg/s}$ bin of the duals and offsets is lower than all AGN, changing the denominator to include pairs (single AGN) of all luminosities can flip this relation.

\section{Evolution of Dual and Offset AGN}
\label{sec:evolution}

\begin{figure*}
\centering
  \includegraphics[width=\textwidth]{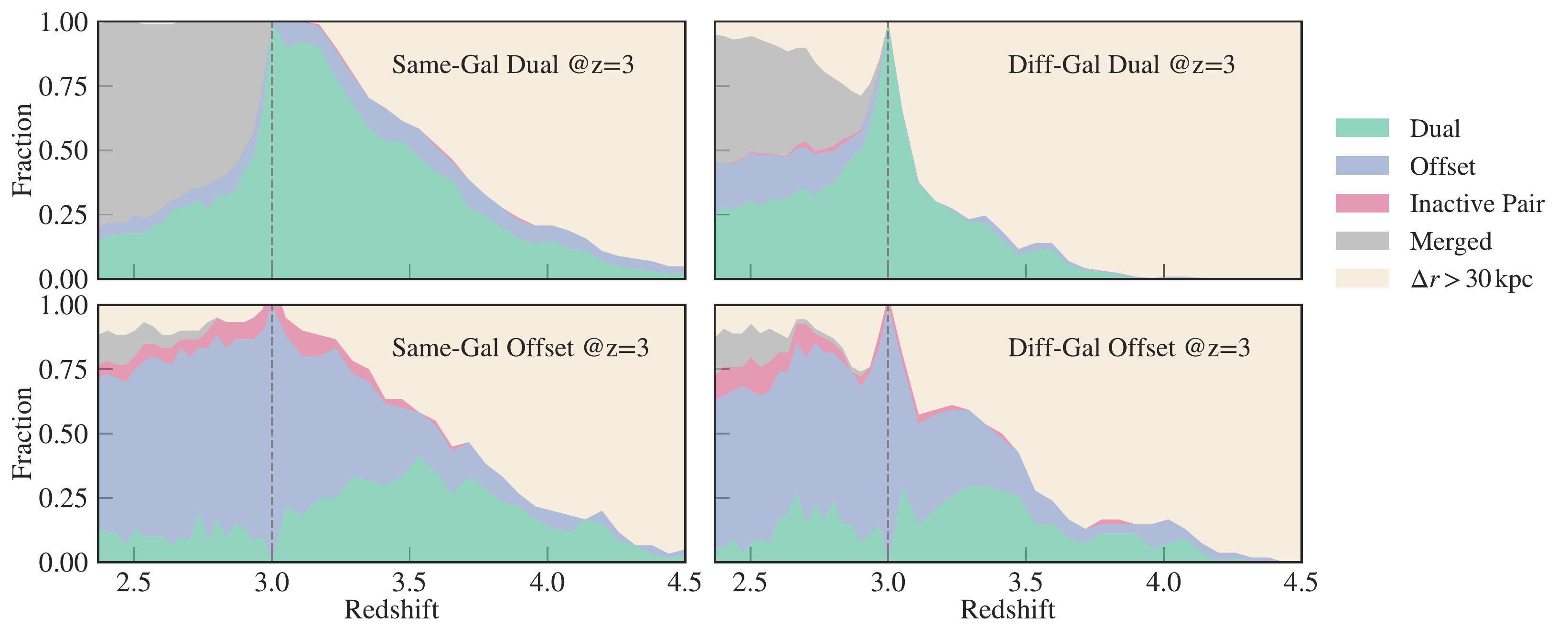}
  \caption{\textit{\textbf{Top:}} the classification of same-galaxy dual AGN (\textit{\textbf{left}}) and different-galaxy dual AGN (\textit{\textbf{right}}) identified at $z=3$, throughout $z=2.4 \sim 4.5$.
  The width of each band corresponds to the fraction of dual AGN falling into each category. We have categorized the pairs into duals (\textit{\textbf{green}}), offsets (\textit{\textbf{purple}}), inactive pairs (\textit{\textbf{pink}}), mergers (\textit{\textbf{grey}}) and non-pairs (\textit{\textbf{beige}}) . Here the dual and offset categories have been previously defined for our work. Inactive pairs are pairs with both MBHs under $L_{\rm bol}<10^{43}\,\mathrm{erg}/s$.
  The merger category refers to simulation mergers, with the merging criterion defined in Section \ref{sec:sim}.
  }
  \label{fig:frac_ev}
\end{figure*}

Up until now, we have focused on dual and offset AGN properties at a fixed redshift of $z=2$, in order to make comparisons with observational properties.
In this section, we will take advantage of the simulation's access to the evolution of the pairs across different redshifts and the evolutionary stage of the pairs.
For this purpose, we will use the 329 dual and 110 offset AGN samples at $z=3$, because we would like to trace the evolution of those pairs both before and after the time of observation.

\subsection{Connections between pairs across different redshifts}

Previous simulations have associated AGN triggering with galaxy mergers, showing that the MBH pairs are observable as duals only during a small fraction of time during the host galaxy merger.
Hence, a fraction of the observable duals at $z=3$ could have been offset AGN or inactive pairs in the past.
Furthermore, some dual AGN may suffer suppression in activation during galaxy mergers, due to the gas outflows and heating induced by the galaxy mergers, and thus become an offset pair at closer separations \citep[e.g.][]{Steinborn2016}.
The future of the duals and offsets is also of great interest, as they are progenitors of MBH merger event.
Here we will look into the past and future of dual and offset pairs, and draw connections between the pair population at different redshifts.

In Figure \ref{fig:frac_ev}, we show the classification of the dual and offsets at $z=3$ through different times before and after the time of observation.
The top panels show the evolution of the same- and different-galaxy duals, where the width of each colored band shows the fraction of $z=3$ duals falling into each category at different times.
We have categorized the pairs into five categories: dual and offset AGN as defined throughout this paper, inactive pairs when the MBHs are separated by $<30\,{\rm kpc}$ but do not fall into the dual and offset categories due to the luminosity threshold (not the mass threshold), mergers when the two MBHs merge in the simulation, and $\Delta r>30{\rm kpc}$ when the two MBHs do not form a pair.
For the same-galaxy duals at $z=3$, the pairs were formed as early as $z=4.5$, and $50\%$ of those duals were observable as duals from $z=3.5$ to $z=3$.
After $z=3$, the same-galaxy duals go through rapid mergers because they are already very close at the time of observation, and also have similar masses compared to offset pairs (so that the dynamical friction time is short).
By $z=2.4$, more than $80\%$ of these duals would merge in the simulation.
We note that mergers in the simulation do not guarantee an MBH coalescence, due to the sub-resolution dynamical friction time \citep[e.g.][]{Chandrasekhar1943,Dosopoulou2017,Pfister2019} and the binary hardening time \citep[e.g.][]{Sesana2010,Vasiliev2015}, as well as the possibility of a three-body scattering.
Nonetheless, in the regime where both MBHs are massive, the delay due to the above mechanisms is expected to be within $\sim {\rm Gyr}$ \citep[see e.g.][]{ChenNianyi2022}.
Finally, about $5\%$ of the same-galaxy duals would fall into the offset AGN category at other redshifts.
They may be observable as duals at $z=3$ only because of the time variability of the AGN activation.

The different-galaxy duals at $z=3$ have come to within $30\,{\rm kpc}$ of each other more recently, with more than $50\%$ forming pairs after $z=3.2$.
The different-galaxy duals can be viewed as the progenitors of three distinct populations when we look at their evolution after $z=3$.
At $z=2.5$, only $\sim 25\%$ of the different-galaxy duals remain to be dual AGN, and most would become small-separation, same-galaxy duals.
Another $25\%$ of the different galaxy duals would evolve into offset AGN, due to the gas disruption of the secondaries during the galaxy close encounters.
Finally, an increasing fraction of the different-galaxy duals would first become same-galaxy duals, with $\sim 50\%$ having merged at $z\sim 2.5$.
Notably, $10\%$ of the different galaxy duals would be separated by more than $30\,{\rm kpc}$ after $z=2.5$, while $\sim 25\%$ are separated to $>30$kpc shortly after $z=3$, before getting closer again.
This is because at $z=3$, we happened to have caught those duals at their pericentric passage, and they will get into larger separations for some time before settling into $\Delta r<30\,{\rm kpc}$ orbits.

Next, we show the history of offset AGN at $z=3$ on the bottom panels of Figure \ref{fig:frac_ev}.
For offset AGN at $z=3$, at least $35\%$ were once dual AGN at $z\sim 3.5$, and have only become offset pairs between $z=3$ and $z=3.5$. 
In fact, if we trace the same-galaxy duals and offset pairs back to $z=3.5$, we see an equal fraction of them were once dual AGN, but these $z=3.5$ duals then quickly parted ways and evolve into duals and offsets at $z=3$.
We will investigate the reasons for the diverging paths of the duals and offsets during this time in later sections.

When we follow these offsets to lower redshifts, we can see that once the pair becomes an offset, it will very likely remain so for a very long period of time, without going through mergers or both MBHs becoming active again.
The stellar stripping of the offset MBH leads to very inefficient orbital decay \citep[also see e.g.][]{Tremmel2018b}, such that the offset stalls at relatively large orbits for up to $>{\rm Gyrs}$.
During this long period of in-fall time, we also see that $\sim 10\%$ of the same-galaxy offset will be dissociated, likely due to the disruption from a third galaxy.
Finally, only around $5\%$ of the offsets will be observable as a dual at a given time, mostly during the pericentric passages when the secondary MBH passes through the high-density regions near the primary AGN.

\subsection{Pair evolution during galaxy mergers: case studies}

\begin{figure*}
\centering
  \includegraphics[width=\textwidth]{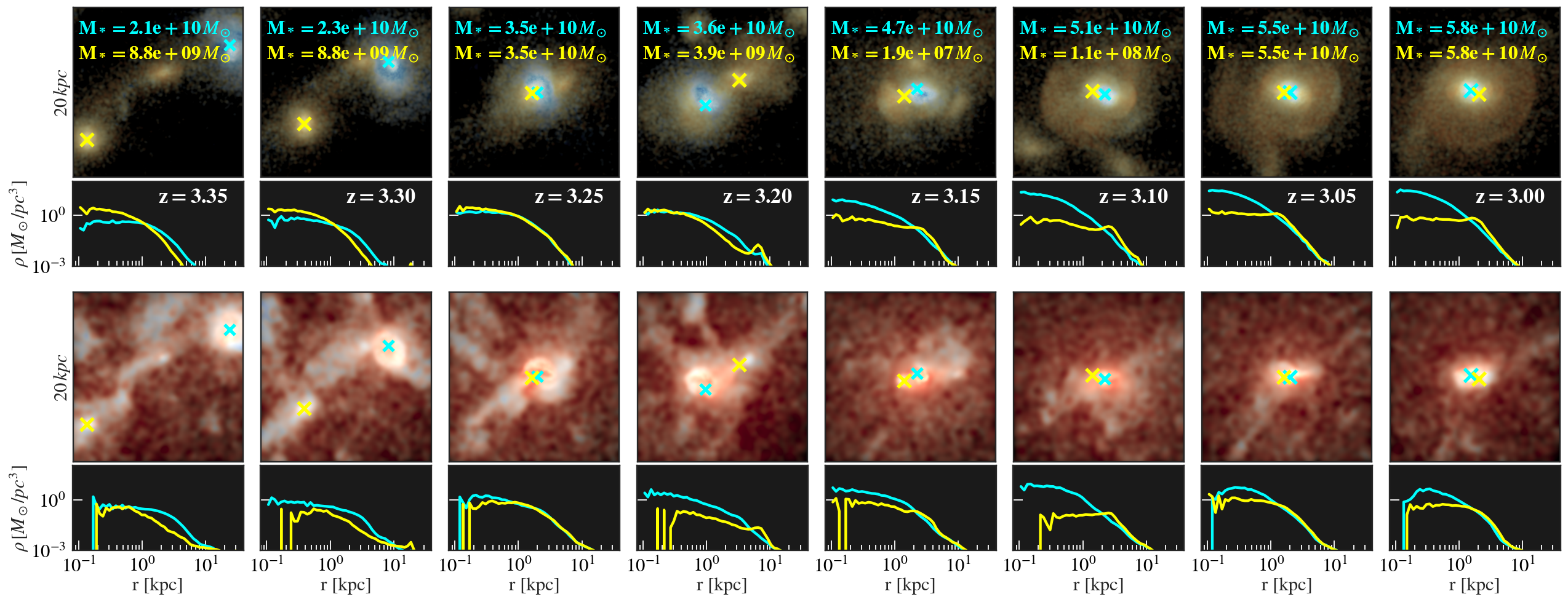}
  \caption{The galaxy (\textit{\textbf{top}}) and gas (\textit{\textbf{bottom}}) surrounding a dual AGN during the galaxy merger. The galaxies are color-coded by the stellar age (warmer colors correspond to older stars), and the gas is color-coded by the gas temperature (warmer colors correspond to higher temperature), with brightness representing the densities for both. The crosses mark the two MBHs.
  The bottom panels show the gas and stellar densities around the MBH with the corresponding color. The host galaxy masses of the two MBHs are marked with the corresponding color. The host galaxies merge between $z=3.1$ and $z=3.05$.}
  \label{fig:dual_ex1}
\end{figure*}

\begin{figure*}
\centering
  \includegraphics[width=\textwidth]{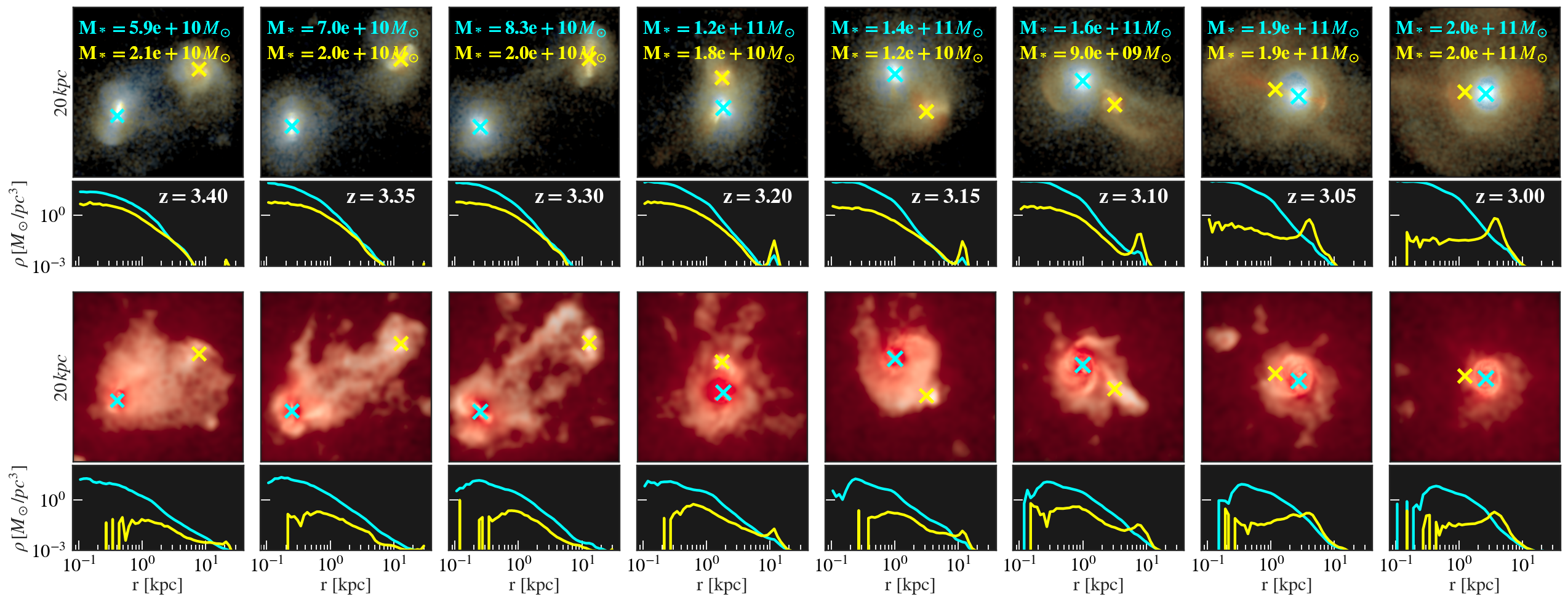}
  \caption{Similar to Figure \ref{fig:dual_ex1}, but for an MBH pair that evolved from a dual AGN to an offset AGN after the galaxy merger.
  The color ranges of the gas and galaxy images are the same as the dual pair for comparison.
  The host galaxies merge between $z=3.1$ and $z=3.05$.
  After $z=3.1$, the secondary galaxy is almost completely disrupted and the secondary MBH becomes an inactive bare MBH.
  We also note that the gas temperature around the offset pair is higher, and we find this to be generally true among offsets.}
  \label{fig:offset_ex1}
\end{figure*}

\begin{figure}
\centering
  \includegraphics[width=0.49\textwidth]{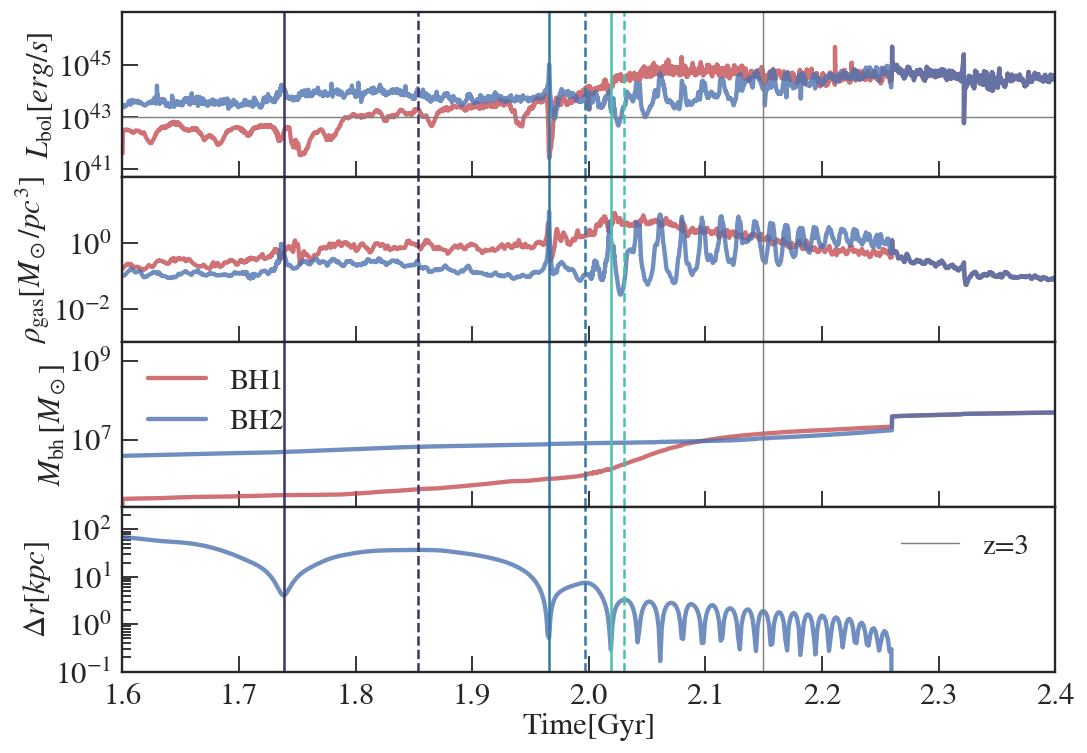}
    \includegraphics[width=0.49\textwidth]{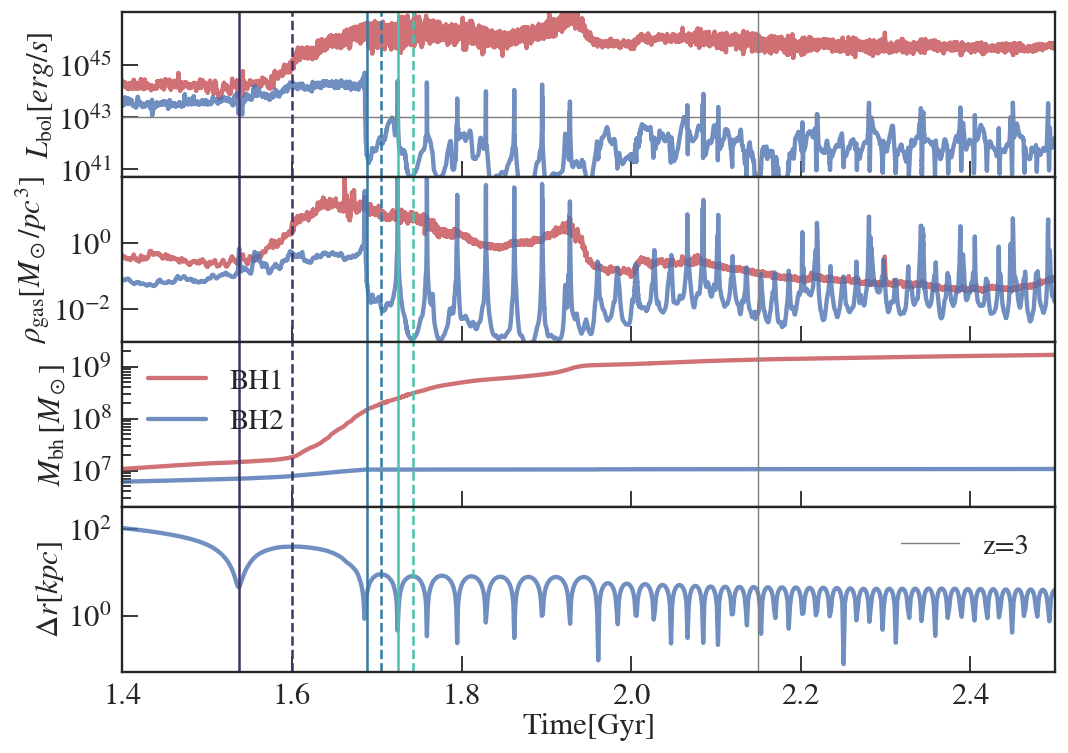}
  \caption{The evolution of a dual (\textit{\textbf{top}}) and an offset pair (\textit{\textbf{bottom}}), where we traced their luminosity (\textit{first panel}), surrounding gas density (\textit{second panel}), masses (\textit{third panels}) and the pair separation (\textit{fourth panel}), throughout the pair formation time. We mark the time of the first (\textit{dark blue}), second (\textit{blue}) and third (\textit{green}) pericentric (\textit{solid}) and apocentric (\textit{dashed}) passages by the vertical lines.}
  \label{fig:curves}
\end{figure}

Previously, we have found that dual and offset AGN are luminous at a closer pair separation, and that the high-mass ratio, different-galaxy duals may be the progenitor of offset AGN at a closer separation.
In this subsection, we will follow some of the $z=3$ duals and offset AGN throughout their formation histories, and investigate their surrounding environment and activation during different stages of the galaxy mergers.

In Figure \ref{fig:dual_ex1}, we show the evolution of the galaxy and gas surrounding the two MBHs in a $z=3$ dual AGN during the pair formation, where the galaxies are color-coded by the stellar age (warmer colors correspond to older stars), and the gas is color-coded by the gas temperature (warmer colors correspond to higher temperature), with brightness representing the densities for both.
The bottom panels show the density profiles of the stars and gas, centered around the two MBHs.
The evolution of the luminosities, gas densities, masses, and the separation between the two MBHs in the same pair are shown in the top panel of Figure \ref{fig:curves}.
For this system, we see that the initially less-massive and inactive MBH (cyan in Figure \ref{fig:dual_ex1} and red in Figure \ref{fig:curves}) was embedded in a galaxy with newly formed stars and a denser cold gas reservoir. 
During the galaxy merger, this less-massive MBH goes through very rapid growth by accreting from its surrounding gas and becomes the brighter AGN at the third pericentric passage.
Then, after about $50\,{\rm Myrs}$, its mass also catches up with the initially more massive MBH (yellow in Figure \ref{fig:dual_ex1} and blue in Figure \ref{fig:curves}).
In the meantime, the central density of its surrounding gas also grows by more than an order of magnitude.

The chosen dual AGN is typical among the few hundred duals, although there is a large variance among the population (see e.g. the distribution shown in Figure \ref{fig:mstar_evolve}).
The evolution of this pair is also in concordance with the cases of idealized galaxy mergers presented by \citet{Callegari2009,Callegari2011} and \citet{Wassenhove2012} in many aspects:
during the galaxy merger, there is a weaker SFR in the initially more massive MBH, while the initially lighter MBH has a higher central SFR during the galaxy merger, building up a dense cusp while outgrowing and disrupting the initially more massive MBH.

In Figure \ref{fig:offset_ex1} and the bottom panels of Figure \ref{fig:curves}, we show the same information but for an offset AGN pair at $z=3$.
Compared with the dual AGN example, both MBHs in this offset pair are more massive, but we note that the mass contrast between the two MBHs in this offset pair before the encounter is actually smaller, indicating that minor mergers are not necessary conditions for forming an offset pair.
The large-scale environment can also play a key role: the gas temperature surrounding the offset pair is hotter, and we find this to be generally true for the majority of offset AGN, compared with the dual AGN.
The hotter ISM/environment can be attributed to these objects being embedded in a more massive halo than a typical dual. 
This is illustrated in Figure \ref{fig:mfof}.

From the luminosity and gas density shown on the bottom panel of Figure \ref{fig:curves}, we can see that between the first and the second pericentric passages, the primary AGN goes through a very rapid phase of gas accretion, when its surrounding gas density increases by two orders of magnitude, and its luminosity also increases by two orders of magnitude.
On the other hand, there is a very clear gas stripping of the gas surrounding the secondary MBH immediately after the secondary pericentric passage (marked by the purple solid lines). 
After this point, the secondary MBH remains inactive for the majority of the time, although occasionally during the pericentric passages of the orbits, the inactive secondary comes very close to the primary, so that its luminosity peaks above $10^{43}\,\mathrm{erg}/s$.

From the pair separation shown in the fourth panel of Figure \ref{fig:curves}, we can also see that another effect of the complete star and gas stripping of the secondary is that the MBH orbit remains large for a very long time: even though the galaxy merger of the offset pair takes place before the dual example shown in the top panel, the MBHs do not merge in the next Gyr.
This is a further illustration of why the offset pairs remain offsets for an extended period of time in Figure \ref{fig:frac_ev}.

\begin{figure}
\centering
  \includegraphics[width=0.49\textwidth]{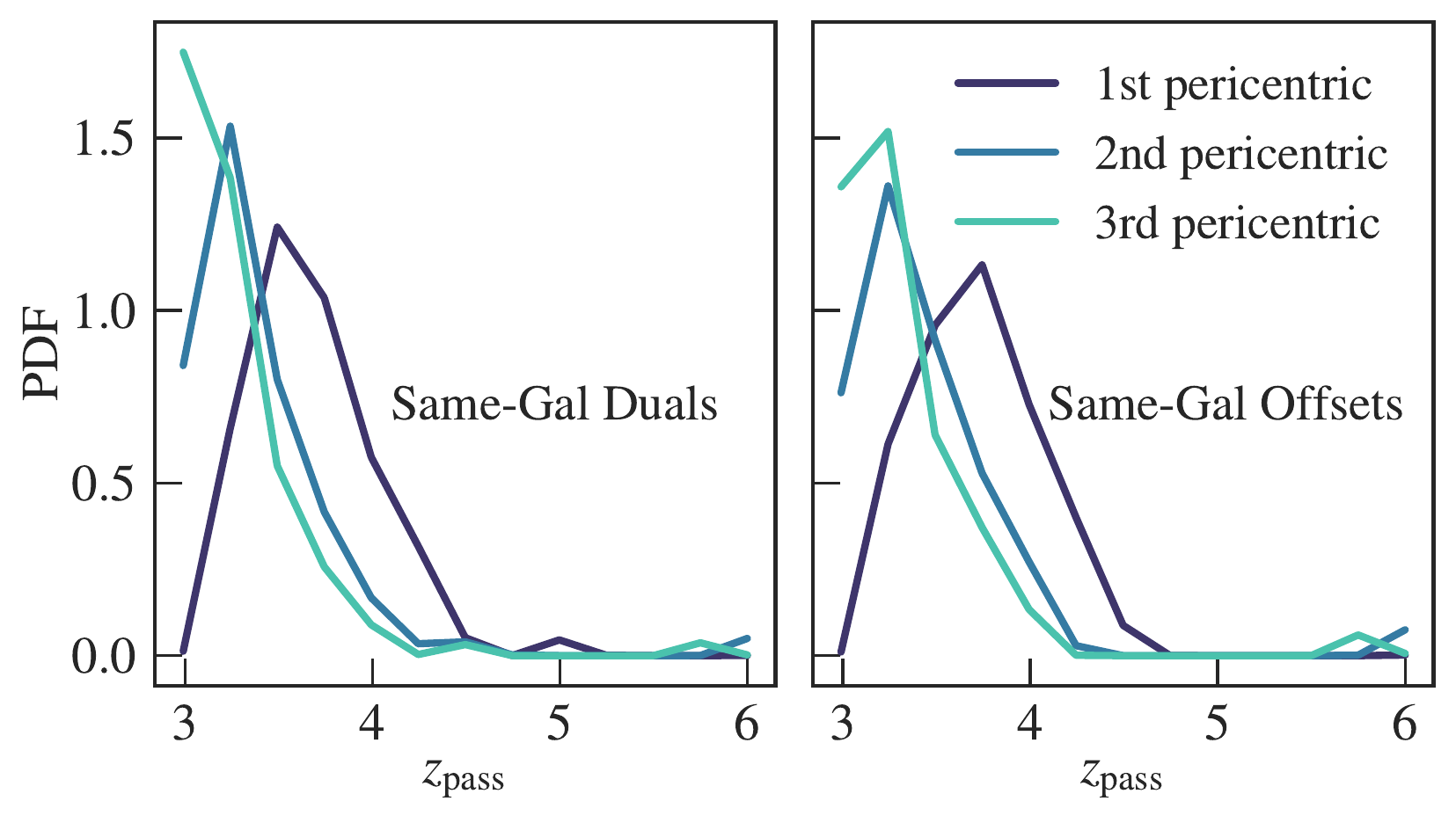}
  \caption{The redshift distribution of the first three pericentric passages between the same-galaxy dual AGN pairs (\textit{\textbf{left}}) and offset AGN pairs (\textit{\textbf{right}}) at $z=3$. Duals and offsets have similar pericentric passage times, but the orbital periods of offsets are shorter, leadning to an earlier third passage (as was illustrated in the cases of Figure \ref{fig:curves}).}
  \label{fig:zpass}
\end{figure}

\begin{figure}
\centering
  \includegraphics[width=0.49\textwidth]{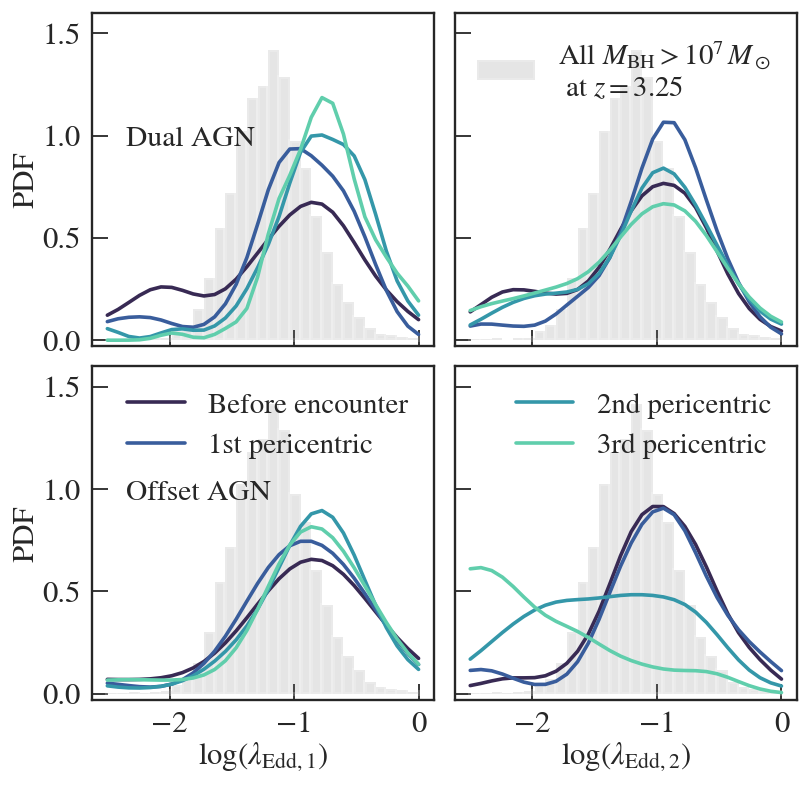}
  \caption{The distributions of the Eddington ratio of the $z=3$ same-galaxy duals (\textit{\textbf{top}}) and offsets (\textit{\textbf{bottom}}), traced back to before the encounter galaxies, and the first, second and third pericentric passages of the MBHs. The \textit{\textbf{left column}} shows the Eddington ratio of the primary AGN, and the \textit{\textbf{right column}} shows the Eddington ratio of the secondary AGN. We also plot the the Eddington ratio of all MBHs with $M_{\rm BH}>10^7\,M_\odot$ at $z=3.25$ (250 Myrs before $z=3$, \textit{grey}) for reference.}
  \label{fig:eddington}
\end{figure}

\begin{figure}
\centering
  \includegraphics[width=0.48\textwidth]{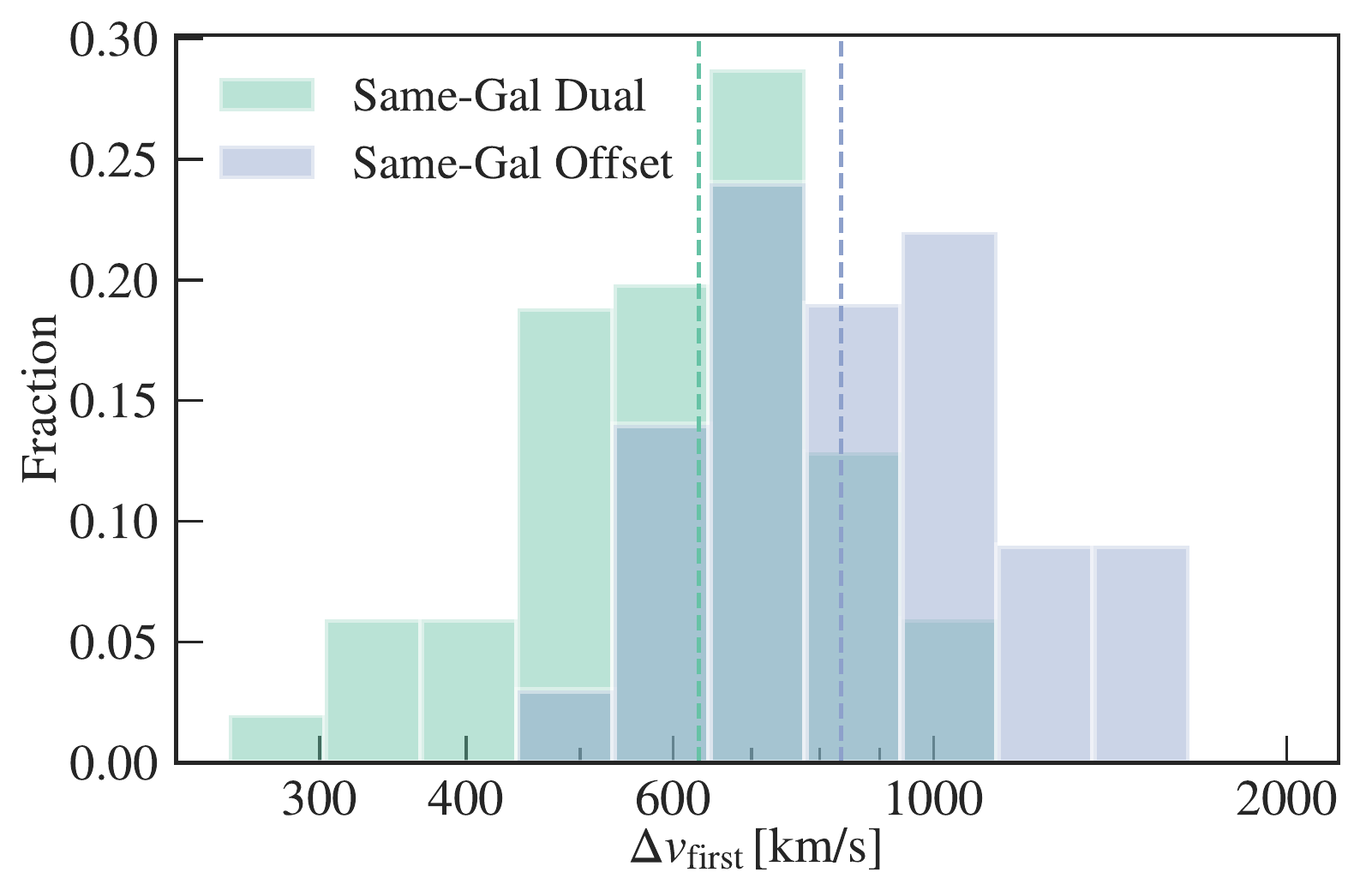}
  \caption{The distribution of the velocity offset between the two MBHs in the $z=3$ same-galaxy duals (\textit{\textbf{green}}) and offsets (\textit{\textbf{purple}}) at the first pericentric passage. The dashed lines show the median of each distribution. The offset AGN generally have larger velocity offsets compared to duals.}
  \label{fig:velocity}
\end{figure}

\subsection{Pair evolution during galaxy mergers: population statistics}
\label{subsec:pair_evol}

\begin{figure*}
\centering
  \includegraphics[width=0.99\textwidth]{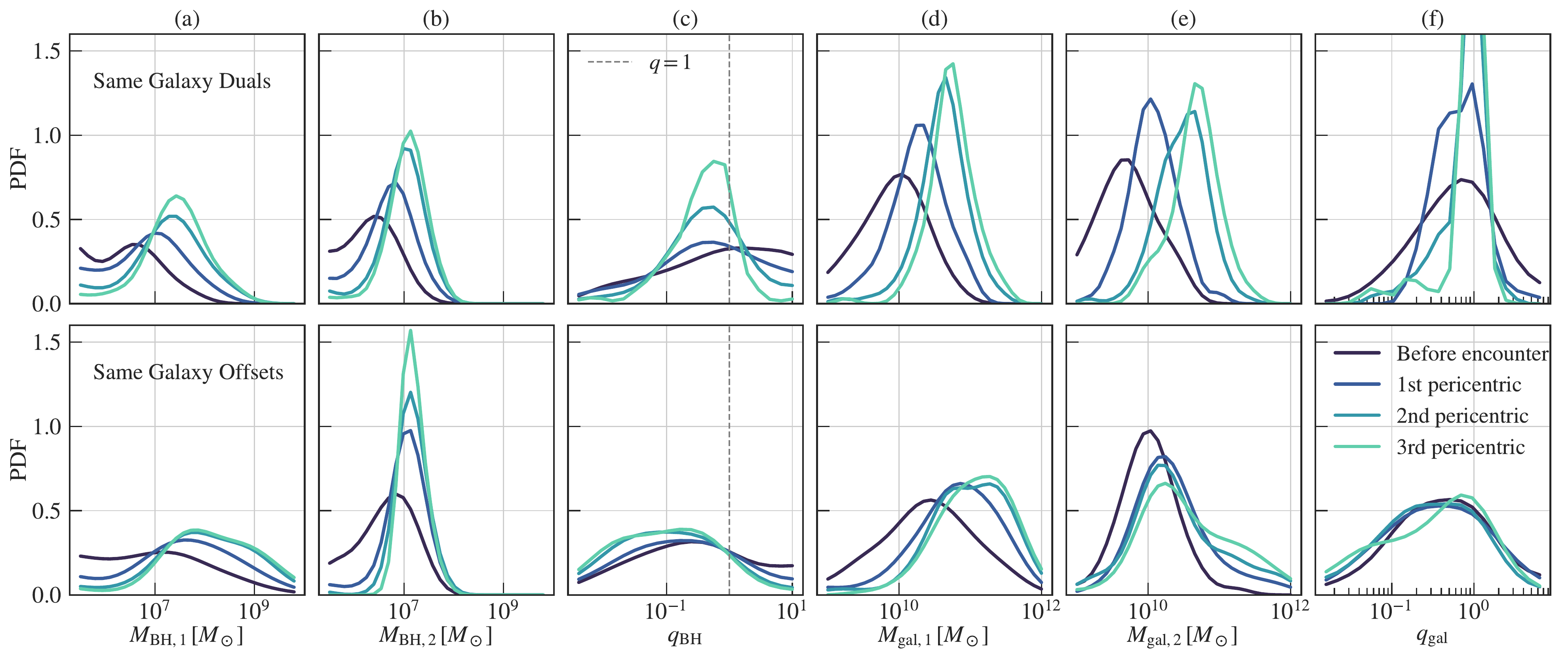}
  \caption{
  \textit{\textbf{(a/b)}}: the mass evolution of the two MBHs in the dual (\textit{\textbf{top}}) and the offset (\textit{\textbf{bottom}}) pairs before the encounter of the host galaxies, and at the first, second and third pericentric passages between the MBH pair. Here BH1 is the more massive MBH at $z=3$, and BH2 is the less massive MBH at $z=3$.
  \textit{\textbf{(c)}}: mass ratio between the two MBHs in the pair.
  \textit{\textbf{(d/e)}}: the stellar mass of the host galaxies for each MBH in the dual and offset pairs. If the galaxies of the two MBHs merged, then we show the mass of the merger remnant.
  \textit{\textbf{(f):}} the evolution of the galaxy mass ratio between the galaxy hosting the secondary MBH and that hosting the primary MBH.}
  \label{fig:mstar_evolve}
\end{figure*}

From the previous section, we have selected two typical cases among the dual and offset AGN at $z=3$ to illustrate the difference in their time evolution.
Now we will apply a similar time-evolution analysis but for the whole dual and offset population.

Similar to the case studies, the point at which the pericentric passage is identified is shown by the vertical lines of the same colors in the fourth panel of Figure \ref{fig:curves}.
We show the redshift distribution of the first three pericentric passages for the same-galaxy duals and offsets in Figure \ref{fig:zpass}.
Note that duals and offsets have similar first-passage redshifts, but the orbital periods of offsets are shorter, leading to an earlier third passage among offsets (as was illustrated in the cases of Figure \ref{fig:curves}).
When comparing the pair properties of duals and offsets, the difference between the time of the third passage should also be noted, while the comparison between the first two passages is fair.
Finally, the time before the encounter is defined to be the first time that the two MBHs are separated by less than the virial radius of the smaller subhalo, such that there has not been any significant dynamical interaction between the gas and stars surrounding the two MBHs.

In Figure \ref{fig:eddington}, we show the Eddington ratio of two MBHs involved in the dual and offset pairs at different stages of the galaxy merger.
The top panels show the evolution of the primary MBH at $z=3$ (but note that they are not necessarily more massive at all stages), and the bottom panel shows the evolution of the secondary MBH at $z=3$.
Here we trace the AGN activation before the galaxy merger and during the first, second, and third pericentric passages of the orbits.

For the dual AGN, both MBHs start off at an Eddington ratio slightly above $0.1$, but there are $20\%$ in each population with Eddington ratios below $0.03$.
During the first pericentric encounter, we see an increase in AGN activity in both MBHs of the dual: a large fraction of the initially inactive tails activated with an Eddington ratio around 0.1.
The divergence between the evolution of the two MBHs happens at the second pericentric passage: the primary AGN becomes more active with Eddington ratios peaking around 0.3, whereas some of the secondaries show a decrease in activity.

The effect of pericentric passages on the offset pairs is more significant, as was shown in the right panels of Figure \ref{fig:eddington}.
From the plot on the left,  we can see that the activation of the primary MBH of the offset steadily increases with each pericentric passage.
The secondary MBHs (which are inactive at $z=3$) start at a similar Eddington ratio as the primary, with an Eddington ratio of $\sim 0.1$, and maintains this Eddington ratio at the first pericentric passage. 
However, we see a sharp decrease in the Eddington ratio at the second pericentric passage, when half of the secondaries now have Eddington ratios below $\sim 0.03$.
At the third passage, the gas stripping is more severe, and the majority of the secondary becomes inactive with Eddington ratios below $0.01$.
This is illustrated earlier in Figure \ref{fig:curves}, where we see a very sudden drop of the secondary MBH's surrounding gas density at exactly the second pericentric passage.

The strong gas stripping and the deactivation of the initially bright secondary among offsets is a result of many factors.
For example, \citet{Callegari2011} and \citet{Wassenhove2012} find that compared to coplanar galaxy mergers, inclined mergers can have less
central star formation in the secondary, leading to disruption at $\sim {\rm kpc}$ separations rather than efficient pairing.
Here we investigate one specific factor that could affect the degree of gas stripping: the velocity difference between the two MBHs at their first pericentric passages.
In Figure \ref{fig:velocity}, we show the velocity difference $\Delta v_{\rm first}$ between the two MBHs in duals and offsets, at their first passage.
Compared with duals, offset pairs have a larger velocity offset at the first encounter.
This could also be a consequence of the deep potential associated with the large host halo of offset pairs.
The high potential energy is transformed into high kinetic energy during the pericentric passages.
Since large ($\Delta v>150\,{\rm km/s}$) velocity offsets between AGN pairs is also important for the pair detection, offset AGNs will more likely satisfy the velocity criterion than duals.

%----------------------------------

In Figure \ref{fig:mstar_evolve}, we show the evolution of the MBH masses and host galaxy masses for the dual and offset pairs, as well as the MBH and galaxy mass ratio between the host galaxies.
Here we denote the more massive MBH at $z=3$ as the primary (BH1) and its host as  ${\rm gal1}$, and the less massive MBH at $z=3$ as the secondary (BH2) and its host as ${\rm gal2}$.

Before the encounter of their host galaxies, MBHs in duals have typical BH masses below $10^7\,M_\odot$, and a subset as low as $M_{\rm BH}<10^6\,M_\odot$.
The offset pairs start off with typically higher masses than duals, with about $50\%$ of offsets with BH masses above $10^7\,M_\odot$ before their host first encounter.
During the interaction of their host galaxies, the primary MBHs, in both the dual and offset pairs, accrete a significant amount of mass, and by the end of the third pericentric passage, the peaks, in the respective BH mass distributions, shift up by an order of magnitude.
This is expected from the high Eddington ratio of the primary dual and offset during this period shown in Figure \ref{fig:eddington}.
As for the secondary MBHs in duals, the BH mass growth is less significant, but most secondaries still reach a mass of $10^7\,M_\odot$ by the second passage. However, after this point, the growth stalls and the masses of the secondaries do not go much beyond $10^7\,M_\odot$.
Notably, $\gtrsim 50\%$ of duals have a mass ratio greater than unity before the host galaxy encounter, meaning that the  initially less massive MBH among duals ends up accreting more mass during the galaxy merger and becomes the primary at $z=3$.

We also see similar trends for the respective host galaxy masses of duals and offsets. 
Before the first encounter, the primary MBHs of the duals reside in galaxies with masses between $10^9\,M_\odot$ and $10^{11}\, M_\odot$.
The secondary galaxy is generally slightly less massive, with distribution peaked near $5\times 10^9\,M_\odot$.
Notably, none of the dual pairs has a galaxy mass above $10^{11}\,M_\odot$ before the encounter.
Now looking at the time evolution of the masses, we see that the primary galaxy mass grows by an order of magnitude during the few hundred Myrs of the orbital passages.
The galaxy mass of the secondary grows even faster, and by the time of the third pericentric passage, the majority of the two hosts have already merged and thus share the same galaxy mass.

The evolution of the mass ratio between the two MBHs in the pairs and between their host galaxies are particularly worth noting.
Observational studies such as \citet{Comerford2015} found that all dual AGN and dual AGN candidates in their sample share the feature that the MBH in the less luminous galaxy always has the highest Eddington ratio.
From the host galaxy mass ratio of our dual AGN sample, we see that although the distribution of the secondary's host galaxies does initially peak at a lower value (consistent with observations), about $30\%$ of the duals have an original galaxy mass ratio above unity.
This means that $30\%$ of the more massive MBH in the duals at $z=3$ starts off residing in the smaller galaxy, which then picks up a lot of its mass during the galaxy merger.
\citet{Steinborn2016} also saw a similar trend in their sample of offset AGN.
Our result is also in agreement with high-resolution hydrodynamical simulations of galaxy mergers, which find that the Eddington rate is higher for the AGN in the less massive of the
two merging galaxies \citep[][]{Capelo2015}. 
\citet{Wassenhove2012} also produced situations where the less
massive black hole accretes at a higher Eddington fraction until the less massive galaxy’s gas is lost to ram
pressure stripping.
The higher mass gain in the less massive galaxies can be happen if the less massive galaxies have a higher gas fractions, or if the gas accretion is more efficient in less massive galaxies due to the their stronger gravitational instabilities during mergers.

% \begin{figure}
% \centering
%   \includegraphics[width=0.48\textwidth]{plots/sfr_evol.pdf}
%   \caption{}
%   \label{fig:sfr_evol}
% \end{figure}

% \begin{figure*}
% \centering
%   \includegraphics[width=0.33\textwidth]{plots/prof_same_dual.pdf}
%     \includegraphics[width=0.33\textwidth]{plots/prof_diff_dual.pdf}
%       \includegraphics[width=0.33\textwidth]{plots/prof_offset.pdf}
%   \caption{Gas (\textit{top}) and stellar (\textit{bottom}) density profiles for the dual and offset AGN. \textit{Left:} Densities profiles around the two AGN in the same galaxy. (need some checking to see why the profiles are the same...) \textit{Middle:} Densities around different-galaxy duals. \textit{Right:} Density profiles for all the offset AGN, regardless of whether the two MBHs reside in the same galaxy.}
%   \label{fig:prof}
% \end{figure*}

% \subsection{Large-scale Orbital Properties}

% \subsection{Gravitational-Wave Signals}
% \begin{figure}
% \centering
%   \includegraphics[width=0.4\textwidth]{plots/zmerge.pdf}
%   \caption{}
%   \label{fig:zmerge}
% \end{figure}

% \begin{figure*}
% \centering
%   \includegraphics[width=0.49\textwidth]{plots/pair_counts.pdf}
%   \includegraphics[width=0.49\textwidth]{plots/fraction_in.pdf}
%   \caption{\textit{Left:} Number of MBH pairs as a function of redshift. 
%   The cuts on mass and luminosities are applied to both MBHs in the pair.
%   \textit{Right:} Fraction of MBHs involved in pairs. The same cuts in mass and luminosities are applied to the MBH and pair selection.}
%   \label{fig:count}
% \end{figure*}

\section{Conclusion}
In this work, we characterize the properties and evolution of dual and offset AGN at $z=2\sim3$
within the \texttt{Astrid} simulation, identified with the canonical $L_{\rm bol,12}>10^{43}\,\mathrm{erg/s}$, $M_{\rm BH,12}>10^7\,M_\odot$ and $\Delta r<30\,{\rm kpc}$ thresolds ($L_{\rm bol,2}<10^{43}\,\mathrm{erg/s}$ for offsets).
At these redshifts, dual and offset AGN pairs are very rare (with a number density of $\sim 10^{-5} \, {\rm cMpc}^{-3}$), but with the large volume of \texttt{Astrid}, we are able to identify a statistically large sample of AGN pairs (1087 duals and 842 offsets at $z=2$, 329 duals and 110 offsets at $z=3$).
Having included an on-the-fly subgrid dynamical friction prescription in \texttt{Astrid}, we can trace the MBH orbits down to the resolution limit of $\sim 3\,{\rm ckpc/h}$ ($\sim 1\,{\rm kpc}$ at $z=2$), and capture tens of dual AGN at (small) kpc separations.

Among the massive ($M_{\rm BH}>10^7\,M_\odot$) and luminous ($L_{\rm bol}>10^{43}\,\mathrm{erg/s}$) AGN at $z=2$, the dual fraction is $3\%$ with separations below $30\,{\rm kpc}$. 
Another $2.2\%$ of the AGN are involved in offset pairs, where the secondary is massive but not luminous (with $L_{\rm bol}<10^{43}\,\mathrm{erg/s}$).
We do not see a strong redshift dependence in the dual fraction from $z=2$ to $z=4$, but the fraction drops below $2\%$ above $z=4$.

Out of the $z=2$ dual AGN, $\sim 50\%$ are AGN within the same galaxy and separated by $<5\,{\rm kpc}$. 
The number of duals increases with decreasing separation, and over half of the same-galaxy duals are found at separations below $2\,{\rm kpc}$.
For the dual AGN residing in different galaxies, we do not see a strong separation dependence of the number of duals.
Among the offset pairs, over $80\%$ are found at separations between $2-10\,{\rm kpc}$.
Offset AGN are rare at smaller or larger separations because their formation involves both strong gas stripping which becomes most effective at $<10\,{\rm kpc}$ scales, and large enough separation between the pair such that the two MBHs are not accreting from the same gas reservoir (which would equalize their luminosities).

The luminosities of both AGN in duals increase by up to an order of magnitude with decreasing separation below $\Delta r=5\,{\rm kpc}$, indicating that observations with high luminosity threshold could bias towards close pairs (if the spatial resolution allows for a detection of those pairs).
Nonetheless, we find that the gas column density of duals also increases with decreasing separation, which adds complication to the detection of the close pairs.
At larger separations, we find indications of an enhancement in the dual luminosities at $\Delta r=10-15\,{\rm kpc}$, similar to the result shown in \cite{Stemo2021}.
We then confirm with the time evolution analysis of the duals AGN that there is an enhanced AGN activity among both AGN in the pair following the first pericentric passage, which could lead to this bump.

% The primary AGN in the offsets are among the most luminous AGN at $z=3$, whereas the inactive MBHs in the offset pairs make up the majority of under-luminous MBHs with $M_{\rm BH}>10^7\,M_\odot$ at $z=3$.
% However, when traced back in time, inactive MBHs in the offset pairs are typically more massive compared with the secondary of dual AGN.
% The mass accretion of the inactive offset is halted during the galaxy mergers.

We find that details in dynamical interaction during galaxy mergers play a crucial role in explaining the emergence of both dual and offset populations.
Compared with the typical Eddington ratio of $\sim 0.025$ among the MBHs with $M_{\rm BH}>10^7\,M_\odot$ in the simulation, both the dual AGN and the active AGN in the offset pairs correspond to BHs with increased level of activity  with Eddington ratios peaking around $0.05$.
Following each pair through the initial stages of galaxy mergers, we find that the AGN activity of the primary BH in both duals and offsets increases with each pericentric encounter of the pair due to the enhanced gas supply brought in by the galaxy merger, whereas the secondaries suffer from various degrees of gas stripping mostly starting from the second pericentric passage and are thus tend to be less active over time.
The secondary MBHs of the offset AGN experience the most severe/complete stripping, and typically remain inactive on orbits with $\Delta r>2\,{\rm kpc}$ for a few hundred Myrs.
For the secondaries among duals, the central stellar bulges are not completely disrupted during the galaxy merger. 
Thus, $>80\%$ these secondaries can go though more efficient orbital decays towards a simulation merger, usually within $500\,{\rm Myrs}$.
The host-galaxy disruption among offsets also implies that dual/offset candidates selected through distinguishable galaxy bulges\citep[e.g.][]{Stemo2021} are more likely dual AGN pairs with one of the AGN hidden.

By further investigating the host galaxies of the dual and offset AGN, we find that MBHs involved in duals and offsets are under-massive relative to their hosts, with an $M_{\rm BH}/M_*$ ratio below the median value of the similar-mass MBHs.
One possible reason is that the triggering of star-formation preludes the phase of high MBH accretion during galaxy mergers \citep[also see e.g.][]{Callegari2011, Wassenhove2012}.
Indeed, we find that the pair-hosting galaxies show an enhanced specific star-formation rate compared with galaxies of similar masses, especially after the merger of the two hosts.

Notably, there is a switch between the primary and the secondary MBH as well as in their host galaxy mass in $\sim 50\%$ of dual AGN during the galaxy merger: the initially less massive MBH embedded in the smaller galaxy ends up becoming the primary AGN shortly after the galaxy merger.
This switch mostly takes place after the second pericentric passages between the two BHs/AGN.
Our finding is in concordance with higher-resolution galaxy merger simulations \citep[e.g.][]{Capelo2015} as well as recent observation results \citep[e.g.][]{Comerford2015}.
This may give rise to a significant population of bright off-center AGN in the smaller companion galaxy.

The large separation, different-galaxy dual AGNs are progenitors to both duals and offset AGN at closer separations.
Whether a large-separation dual evolve into a close-separation dual or offset depends largely on the level of gas and star stripping during the first three pericentric passages.
The secondary in offset AGN started with a similar level of activation as the secondary of dual AGN, and are even more massive before the galaxy mergers, but they show a sudden decrease in the AGN activity by the third pericentric passage.
One reason for the more severe gas stripping among offset pairs is that they preferentially resides in some of the most massive halos and the deepest gravitational potential.
The velocity difference between the offset pairs at the first pericentric passage is higher compared with the duals.
Other factors such as the rotation of the galaxies relative to the orbit, and the angle of the initial galaxy merger also play important roles, but we do not explicitly quantify these effect from our samples.

In this work, we do not explicitly separate out the multiple AGN systems from the dual and offets as in e.g. \citet{Volonteri2021}.
Nonetheless, we have checked that the fraction of multiple AGN systems make up $<10\%$ of all pairs and do not have a large impact on the overall statistics.
We defer further study of triples and quadruple systems in \texttt{Astrid} to upcoming work.

\section*{Acknowledgements}
\texttt{Astrid} was run on the Frontera facility at the Texas Advanced Computing Center.

TDM and RACC acknowledge funding from 
the NSF AI Institute: Physics of the Future, NSF PHY-2020295, 
NASA ATP NNX17AK56G, and NASA ATP 80NSSC18K101. TDM acknowledges additional support from  NSF ACI-1614853, NSF AST-1616168, NASA ATP 19-ATP19-0084, and NASA ATP 80NSSC20K0519, and RACC from NSF AST-1909193.
YN acknowledges support from McWilliams Graduate Fellowship.
SB was supported by NSF grant AST-1817256. 
AMH is supported by the McWilliams Postdoctoral Fellowship. MT is supported by an NSF Astronomy and Astrophysics Postdoctoral Fellowship under award AST2001810.

%%%%%%%%%%%%%%%%%%%%%%%%%%%%%%%%%%%%%%%%%%%%%%%%%%
\section*{Data Availability}

The code to reproduce the simulation is available at \url{https://github.com/MP-Gadget/MP-Gadget}, and continues to be developed. Text file forms of the data presented here as well as scripts to generate the figures are available. The dual and offset catalogs including the MBH information and the host galaxy properties are available upon request.

\bibliographystyle{mnras}
\bibliography{main.bib}

\end{document}